\documentclass[aoas]{imsart}

\RequirePackage{amsthm,amsmath,amsfonts,amssymb}
\RequirePackage[authoryear]{natbib}
\RequirePackage[colorlinks,citecolor=blue,urlcolor=blue]{hyperref}
\RequirePackage{graphicx}
\RequirePackage{multirow}
\RequirePackage{mathrsfs}
\RequirePackage{booktabs}
\RequirePackage{listings}
\RequirePackage{bm}
\RequirePackage{float}
\RequirePackage{threeparttable}

\startlocaldefs
\theoremstyle{plain}

\theoremstyle{remark}

\endlocaldefs

\begin{document}

\begin{frontmatter}
\title{Changepoint Detection in Categorical Time Series with Application to Daily Total Cloud Cover in Canada}
%\title{A sample article title with some additional note\thanksref{t1}}
\runtitle{Changepoint detection in categorical time series}
%\thankstext{T1}{A sample additional note to the title.}

\begin{aug}
%%%%%%%%%%%%%%%%%%%%%%%%%%%%%%%%%%%%%%%%%%%%%%%
%% Only one address is permitted per author. %%
%% Only division, organization and e-mail is %%
%% included in the address.                  %%
%% Additional information can be included in %%
\author[A]{\fnms{Mo}~\snm{Li}\ead[label=e1]{mo.li@louisiana.edu}},
\author[B]{\fnms{QiQi}~\snm{Lu}\ead[label=e2]{qlu2@vcu.edu}}
\and
\author[C]{\fnms{Xiaolan L.}~\snm{Wang}\ead[label=e3]{xiaolan.wang@canada.ca}}
%%%%%%%%%%%%%%%%%%%%%%%%%%%%%%%%%%%%%%%%%%%%%%
%% Addresses                                %%
%%%%%%%%%%%%%%%%%%%%%%%%%%%%%%%%%%%%%%%%%%%%%%
\address[A]{Department of Mathematics,
University of Louisiana at Lafayette \printead[presep={,\ }]{e1}}

\address[B]{Department of Statistical Sciences and Operations Research, 
Virginia Commonwealth University \printead[presep={,\ }]{e2}}

\address[C]{Climate Research Division, Science and Technology Branch, 
Environment and Climate Change Canada \printead[presep={,\ }]{e3}}

\end{aug}

\begin{abstract}
Changepoints are essential for homogenizing categorical time series and analyzing their trends and variations. The original total cloud cover in Canada was recorded hourly in tenths (or eighths), exhibiting inherent seasonality and serial correlation. \cite{lu2012extended} introduced an extended cumulative logit model to detect shifts in the annual frequencies of cloud cover conditions. While annual aggregation mitigates seasonality and serial correlation, it shortens the time series and may lead to overdispersion.
This article introduces a marginalized transition model to detect a single changepoint in periodic and serially correlated categorical time series. The model captures serial dependence using a first-order Markov chain and enables category-specific changepoint specification. To enhance computational efficiency, we develop a new parameter estimation procedure for obtaining maximum likelihood estimates. A maximally selected likelihood ratio test statistic is then proposed to test for sudden changes in categorical time series, and the method is illustrated using daily total cloud cover observations recorded at 9 a.m. and 3 p.m. at Fort St.~John Airport, British Columbia, Canada.

\end{abstract}


\end{frontmatter}
%%%%%%%%%%%%%%%%%%%%%%%%%%%%%%%%%%%%%%%%%%%%%%
%% Please use \tableofcontents for articles %%
%% with 50 pages and more                   %%
%%%%%%%%%%%%%%%%%%%%%%%%%%%%%%%%%%%%%%%%%%%%%%
%\tableofcontents

\section{Introduction}\label{MTMIntro}

In numerous climate studies, the analysis of changepoints in non-stationary time series is crucial, as climate series frequently exhibit temporal discontinuities over the data recording period. A changepoint refers to a time point at which the statistical properties of the series undergo an abrupt change. Such change could manifest as a shift in the mean, variance, or correlation structure of the underlying distribution. Changepoints in climate time series can occur for various reasons, often associated with changes in the instrumentation/observer, observation environment, and/or procedure, such as station/instrument relocation and changes in observation time/frequency.
%due to natural climate variability or anthropogenic influences, such as station relocations or instrumentation changes. 
In practice, users of climate data are advised to integrate all available metadata and employ suitable statistical methods to detect artificial changepoints and adjust the data series through homogenization techniques \citep{jovanovic2011high}.

Sky-cloudiness condition (cloud cover) is important for climate studies and understanding the Earth's energy balance. In Canada, visual observations of cloud cover are observed hourly in tenths of the sky dome, resulting in 11 categories (0 for clear sky, 1 for one-tenth of the sky dome covered by clouds, $\ldots$, and 10 for completely overcast). These categories demonstrate a natural ordering, representing the progression from clear sky to overcast based on the portion of the sky dome covered by clouds. Additionally, cloud cover series often show a long-term trend \citep{milewska2004baseline} and large observational uncertainty. By summing up the hourly frequencies for each category into the corresponding annual categorical frequencies of sky-cloudiness conditions and assuming independence between years, \cite{lu2012extended} applied an extended cumulative logit model. This model considers ordinal categories, accounts for temporal trends in category probabilities, and addresses the overdispersion problem (where the variance is larger than the logit model variance) to detect a level shift in annual cloud cover frequencies. However, aggregating hourly cloud cover series annually would significantly shorten the series and lead to overdispersion in count data \citep{corsini2022dealing}, even though it may reduce seasonality and autocorrelation. Furthermore, while the independence assumption provides a reasonable first-order approximation for annual data, it may not hold for daily cloud cover series.  \cite{lund2007changepoint} and \cite{gallagher2022autocovariance} showed that changepoint inferences could be significantly inaccurate when autocorrelation is ignored. As such, changepoint detection in daily cloud cover time series faces substantial challenges, including seasonality, autocorrelation, and non-Gaussian distribution. 

Considerable research has been conducted on changepoint detection for independent categorical data, including multinomial time series \citep{wolfe1990changepoint, braun2000multiple, robbins2011changepoints}, and ordinal time series \citep{lu2012extended, wang2019latent, lam2020variable}. However, there is limited research on changepoint detection for correlated categorical time series. To handle autocorrelation in categorical time series, past response outcomes together with other explanatory variables were incorporated as covariates in \cite{gombay2017retrospective} to detect changes in the coefficients of a multinomial logistic regression model using the efficient score vector of the partial likelihood function. \cite{li2021structural} extended the work of \cite{gombay2017retrospective} to an ordinal time series by fitting a cumulative logistic regression model. While this conditional model is easy to specify, it generally cannot be reformulated in terms of the marginal mean structure, presenting challenges in interpreting model effects such as changepoints. Moreover, the trend component lacks robustness when conditioned on different sets of past responses \citep{fahrmeir2013multivariate}. Recently, \cite{li2022changepoint} proposed adapted cumulative sums (CUSUM) tests for detecting a mean shift in autoregressive latent Gaussian process, which produces categories through clipping techniques. Although the autoregressive ordered probit (AOP) model is easily specified and interpreted for changepoint analysis, its design for changepoint detection has an inherent limitation. The changepoint in the mean function of the continuous latent process affects all categories simultaneously, shifting the time series to higher or lower categories depending on whether the mean shift is positive or negative. When only a subset of categories experiences probability changes while others remain unchanged, or when the direction of changes alternates across categories, the detection power declines significantly, increasing the likelihood of false negatives.
 
 % {\color{red} As the mean changepoint effect is specified through the continuous latent process and influences all categories simultaneously, the approach proposed by \cite{li2022changepoint} primarily focuses on scenarios where the probabilities of all categories change when a changepoint is present, and it does not accommodate the category-specific probability changes. For example, in an ordinal time series with five categories, the probabilities of different categories may change in alternating directions at the changepoint, with the probabilities of categories one, three, and five increasing while those of categories two and four decrease. Additionally, in cases where only a subset of categories experiences probability changes while others remain unchanged or when the direction of probability changes alternates with category order, the test's power declines and the likelihood of false negatives increases.}

The objective of this article is to develop a likelihood ratio-based changepoint detection method for ordinal time series using a marginalized transition model (MTM) \citep{heagerty2002marginalized, lee2007class} that accommodates seasonality and autocorrelation. Specifically, the seasonal effect, temporal trends in category probabilities, and a mean shift could be described in the marginal mean model with the cumulative logit link. The serial correlation is modeled separately via a first-order Markov chain. The MTM is more appropriate if the primary interest lies in category-specific covariate effects (e.g., changepoints) on the response time series. Our method demonstrates substantially greater detection power and practical advantages over existing changepoint tests, especially when categorical probabilities change in varying magnitudes and directions at the changepoint. % The changing effects can be reasonably applied to all categories or selectively to specific categories. 
Furthermore, the interpretation of regression coefficients is invariant to dependence specification for the marginal models. We implement our method using the R and Rcpp computing environment with the code scripts freely available at \url{https://github.com/mli171/MTMAMOC}.

The rest of this article is organized as follows. Section~\ref{data} presents an overview of the hourly total cloud cover time series at Fort St.~John Airport in Canada, which is used in this study. Section~\ref{MTMModel} introduces the marginalized transition model that combines the cumulative logit model for marginal means and the dependence for transition probabilities. A new sequential parameter estimation method is then proposed to efficiently compute the parameter estimates. Section~\ref{LambdaMax} develops a likelihood ratio-type test for detecting a single changepoint in correlated ordinal categorical series. %{\color{red} Comparing with the methods used in \cite{lu2012extended} and \cite{li2022changepoint}} 
The performance of the test statistic is also evaluated in Section~\ref{LambdaMax} through extensive simulations. The proposed approach is further illustrated with a data analysis of the Fort St.~John Airport series in Section~\ref{Application}. Finally, Section~\ref{conclusion} concludes with a brief discussion.

%%%%%%%%%%%%%%%%%%%%%%%%%%%%%%%%%%%%%%%%%%%%%%%%%%%%
%                     Data                         %
%%%%%%%%%%%%%%%%%%%%%%%%%%%%%%%%%%%%%%%%%%%%%%%%%%%%

\section{Data}\label{data}

The hourly total cloud cover time series used in this study was sampled at Fort St.~John Airport, British Columbia, Canada (Climate ID: 1183000) from 1953 to 2012. It is accessible through the National Climate Archives of the Meteorological Service of Canada. The site is located at an elevation of 695 meters above sea level, with a latitude of $56.24 ^\circ$ N and a longitude of $120.74^\circ$ W. The annual aggregation of this series was analyzed in \cite{lu2012extended}. 

Total cloud cover (TCC) is defined as the fraction of the sky dome covered by all visible clouds. Hourly total cloud cover was visually measured in tenths by trained observers at Fort St.~John Airport beginning in the early 1950s \citep{FortSt1953}. These hourly observations have been archived electronically since 1953 \citep{milewska2004baseline}. Recently, the hourly cloudiness in Canada has been measured in eighths instead of tenths \citep{EnvironmentCanada2019}, which is widely used in many countries such as the United States \citep{dai2006recent} and Australia \citep{jovanovic2011high}. These cloud data are subjective measures of sky cover with large variability and potential inhomogeneities induced by changes in observers, instruments, or observing practice. The installation of Automated Surface Observation Systems (ASOS) during the mid-1990s introduced discontinuities in hourly cloud observations in Canada and the United States, as they measure cloud amounts differently from human observers  \citep{milewska2004baseline, dai2006recent}. Due to these factors and the presence of missing values, hourly observations from 1965 to 1994 are utilized in this study, following the approach of \cite{lu2012extended}. In \cite{lu2012extended}, a changepoint in 1978 was detected. Available metadata (station history information) indicates that the weather office at Fort St. John Airport was relocated on November 7, 1979. This relocation coincided with changes in ownership and station type, suggesting a potential change in observers as well \citep{lu2012extended}. 

Despite the inhomogeneities, surface-observed sky-cloudiness conditions still provide the longest historical record available for studying the long-term trend in cloud cover \citep{dai2006recent,  jovanovic2011high}. Numerous studies have analyzed annual or monthly total cloud cover series by aggregating raw hourly observations \citep{sun2001recent, dai2006recent, free2013time, sun2015variability, lu2012extended}. Given the inherent variability and subjectivity of daily TCC, \cite{jovanovic2011high} averaged daily TCC series at 9 a.m. and 3 p.m. local standard time to develop homogenized monthly TCC datasets for 9 a.m. and 3 p.m. Additionally, daytime TCC series were created by simply averaging the monthly 9 a.m. and 3 p.m. series. Similar to \cite{jovanovic2011high}, this paper focuses on the daily total cloud cover at 9 a.m. and 3 p.m. without aggregating it into monthly TCC series. The daily seasonality and autocorrelation are described in the MTM model, along with the temporal trends, which have been extensively reported in various studies \citep{groisman2004contemporary, dai2006recent,milewska2004baseline,sun2000cloudiness, sun2001recent,kaiser1998analysis, kaiser2000decreasing,kruger2007trends,maugeri2001trends,auer2007histalp}.  Furthermore, \cite{lu2012extended} pointed out that changepoint detection in cloud cover series is often confounded by the presence of temporal trends. Identifying and adjusting for changepoints (inhomogeneities) is essential for accurately assessing long-term trends in sky-cloudiness conditions. 

According to the recent official Manual of Surface Weather Observation Standards (MANOBS) from 2013 to 2023, sky cloudiness conditions are classified into five categories based on hourly observations in tenths: sky clear (0), few (1/10-3/10), scattered (4/10 - 5/10), broken (6/10 - 9/10), and overcast (10/10) \citep{EnvironmentCanada2015}. Alternatively, when using hourly observations in eighths, the categories are sky clear (0), few (1/8-2/8), scattered (3/8 - 4/8), broken (5/8 - 7/8), and overcast (8/8) \citep{EnvironmentCanada2019}. Since the hourly TCC at Fort St. John Airport was observed in tenths, the above five categories are employed as outlined in \cite{EnvironmentCanada2015}. Figures~\ref{fig1} and S1 in the supplementary materials display the hourly TCC series observed at 9 a.m. and 3 p.m., respectively, at Fort St. John Airport from 1965 to 1994, categorized into five sky-cloudiness conditions.

\begin{figure*}[h!]
\centerline{\includegraphics[scale=0.33]{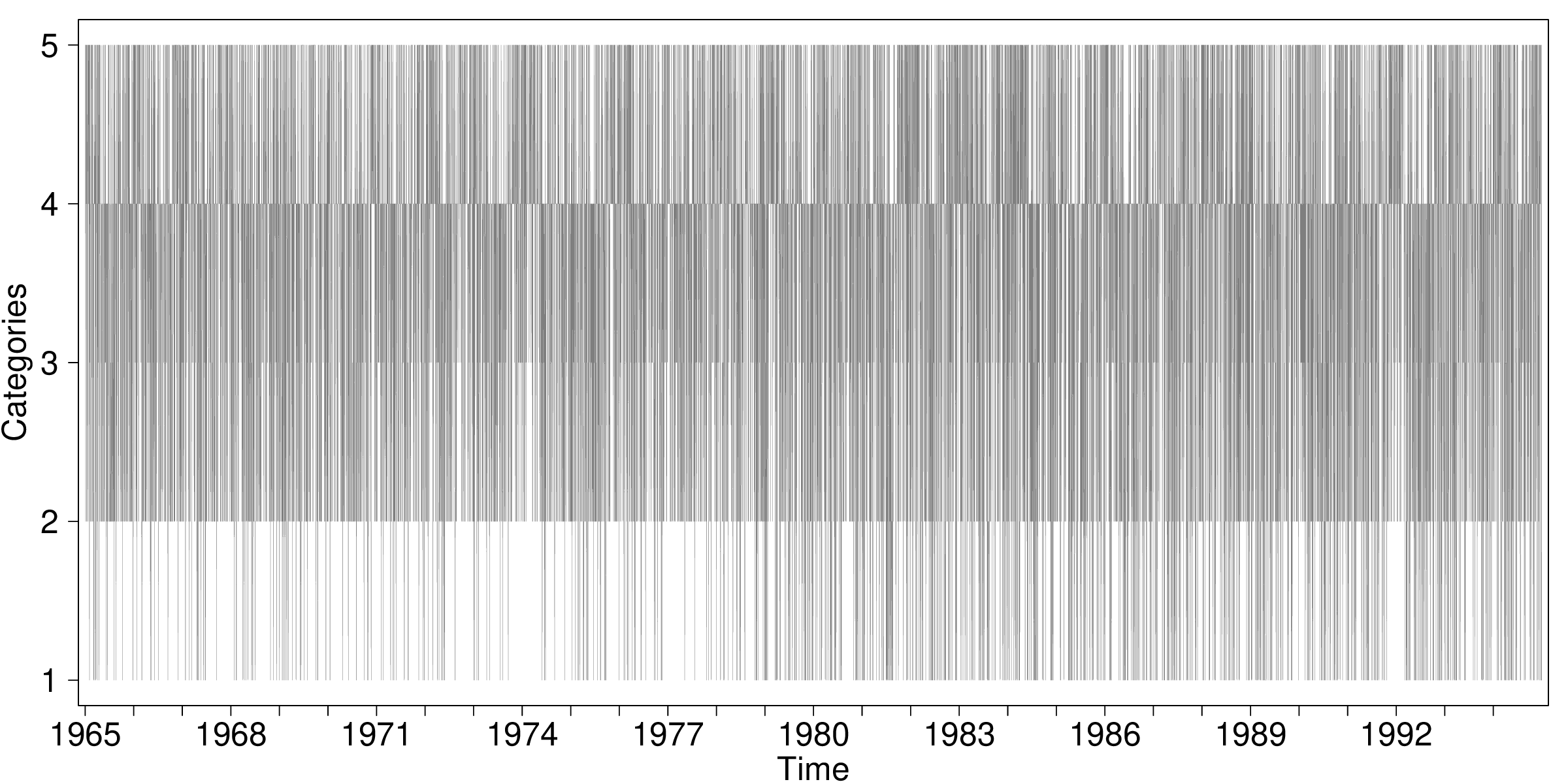}}
\caption{Daily sky-cloudiness conditions observed at 9 a.m. with $K=5$ categories at Fort St. John Airport from 1965 to 1994. \label{fig1}}
\end{figure*}

%%%%%%%%%%%%%%%%%%%%%%%%%%%%%%%%%%%%%%%%%%%%%%%%%%%%
%         Marginalized Transition Model            %
%%%%%%%%%%%%%%%%%%%%%%%%%%%%%%%%%%%%%%%%%%%%%%%%%%%%

\section{The Marginalized Transition Model}\label{MTMModel}
This section describes the MTM for correlated ordinal categorical time series, which consists of two components: the marginal mean model and the dependence model. The MTM originates from the first-order Markov chain model proposed by \citet{azzalini1994logistic} for serially correlated binary data and was later classified by \citet{heagerty2002marginalized} to support likelihood-based marginal generalized linear regression. Subsequently, \citet{lee2007class} extended the model to longitudinal categorical data with ordinal responses by characterizing the marginal means using a cumulative logit model and specifying serial correlation through a Markov transition structure. Similar to \citet{lee2007class}, we extend the cumulative logit model of \citet{lu2012extended} into a marginalized transition model to fit our ordinal time series.

\subsection{Model Formulation \label{Modelform}}
Consider a $K$-category ordinal time series $\{Y_{t}: t=1, \ldots, n\}$. $Y_{t}=k$ indicates that the variable $Y_{t}$ falls into category $k = 1, \ldots, K$ at the time $t$. The category probability vector is $\bm{\pi}_{t}=(\pi_{t,1}, \ldots, \pi_{t,K})'$ with $\pi_{t,k}=\text{P}(Y_{t}=k)$ and $\sum_{k=1}^{K}\pi_{t,k}=1$. Let $\gamma_{t,k}$ denote the cumulative probability that $Y_{t}\leq k$,
\begin{equation*}\label{eqgammdef}
	\gamma_{t,k}=\text{P}(Y_{t}\leq k)=\pi_{t,1}+\pi_{t,2}+\ldots +\pi_{t,k}, \quad k=1, \ldots, K-1.
\end{equation*}
The corresponding cumulative logits are defined as
\begin{equation}\label{eqlogitgamm}
\text{logit}(\gamma_{t,k})=\text{log}\bigg(\frac{\gamma_{t,k}}{1-\gamma_{t,k}}\bigg)=\text{log}\bigg(\frac{\pi_{t,1}+\ldots+\pi_{t,k}}{\pi_{t,k+1}+\ldots+\pi_{t,K}}\bigg), \quad k=1, \ldots, K-1.
\end{equation}
Let $\eta_{t,k}=\text{logit}(\gamma_{t,k})$ and $\tau \in \{1, \ldots, n-1\}$ be an admissible changepoint. The marginal means are specified via the cumulative logit model,
\begin{equation}\label{eqeta}
\eta_{t,k}=\alpha_{k}+s_{t,k}+\beta_{k}\frac{t}{n}+\Delta_{k}I_{[t >\tau]}, \quad k=1,\ldots,K-1;t=1,\ldots,n.
\end{equation}
For each cumulative category $k=1, \ldots, K-1$, $\alpha_{k}$ is the fixed intercept, $\beta_{k}$ is the scaled linear trend parameter, and $\Delta_{k}$ is the category-specific changepoint parameter. The $s_{t,k}$ describes the seasonal effect that is assumed to obey the first-order Fourier expansion,
\begin{equation}\label{eqalpha}
s_{t,k}=B_{k}\text{cos}\bigg(\frac{2\pi t}{T}\bigg)+D_{k}\text{sin}\bigg(\frac{2\pi t}{T}\bigg), \quad k=1,\ldots,K-1,
\end{equation}
where $T$ is the known period of time series. Higher order Fourier expansions or other seasonal cycles could be adapted by adding more sinusoidal pairs in \eqref{eqalpha}. The cumulative probabilities $\gamma_{t,k}$ are obtained as 
\begin{equation}\label{eqgammCalc}
\gamma_{t,k}=\frac{\text{exp}(\eta_{t,k})}{1+\text{exp}(\eta_{t,k})},
\end{equation}
and the categorical probabilities $\pi_{t,k}$ can then be calculated from $\gamma_{t,k}$'s as 
\begin{equation}\label{marginalprob}
    \pi_{t,k} = \begin{cases}
        \gamma_{t,1} & \text{if} \ k=1;  \\
        \gamma_{t,k}-\gamma_{t,k-1} & \text{if} \ k>1.
    \end{cases}
\end{equation}
% $\pi_{t,1}=\gamma_{t,1}$ and $\pi_{t,k}=\gamma_{t,k}-\gamma_{t,k-1}$ for $k=2,\ldots,K$. 
Here, $\gamma_{t,K}=1$. 
% It is worth noting that the model constraint $\eta_{t,1} < \eta_{t,2}< \ldots < \eta_{t, K-1}$ considerably complicates the parameter estimation.  

Next, we develop the dependence model as follows. Assume that $\{Y_{t}\}$ forms a first-order Markov chain with the one-step-ahead transition probability matrix $\mathbb{P}(t)=(p_{t,jk})^{K}_{j,k=1}$ over the $K$ states (categories). In the first-order Markov model, the current response variable depends on the history only through the immediate previous response. Hence, the transition probability of moving from category $j$ at time $t-1$ to category $k$ at time $t$ can be described as $p_{t,jk}=P(Y_{t}=k|Y_{t-1}=j)$ for $1\leq j,k\leq K$ and $\sum_{k=1}^{K}p_{t,jk}=1$. For example, $p_{t,11}=P(Y_{t}=1|Y_{t-1}=1)$ is the probability that the sky is clear at the current time, given that the sky was clear in the previous time. The first-order model assumes that $Y_{t}$ is conditionally independent of $Y_{t-2}, Y_{t-3}, \ldots,$ given $Y_{t-1}$. All dependence in the model is included by the nonhomogeneous K-state Markov chain. Let $P(Y_{t}=k|Y_{t-1})=\pi_{t,k}^{t-1}$ denote the conditional probability. Now we specify the dependence model as
\begin{equation}\label{eqDep}
\text{log}\bigg(\frac{\pi_{t,k}^{t-1}}{\pi_{t,K}^{t-1}}\bigg)=\delta_{t,k}+\sum_{j=1}^{K-1}\xi_{kj}Y_{t-1,j}, \quad k=1,\ldots, K-1; t=2, \ldots, n,
\end{equation}
where for $j=1, \ldots, K-1$, $Y_{t-1,j}=1$ if $Y_{t-1}=j$ and $Y_{t-1,j}=0$, otherwise. The previous response vector $(Y_{t-1,1}, ..., Y_{t-1,K})'$ follows a multinomial distribution with one total event at time $t$. As mentioned in \cite{skrondal2007redundant}, overdispersion is not an issue in this situation. However, for aggregated annual frequencies, the extra-multinomial variation can be accounted for by allowing an overdispersion parameter in the model fitting \citep{lu2012extended}.

The parameter $\xi_{kj}$ in \eqref{eqDep} describes the temporal dependency of $Y_t=k$ on $Y_{t-1}=j$ within the categorical time series. 
The time-category-specific parameters $\delta_{t,k}$'s are implicitly determined by the marginal mean regression parameters and $\xi_{kj}$'s through the relation
\begin{equation*}\label{eqMargDepLink}
\pi_{t,k}=\sum_{j=1}^{K}p_{t,jk}\pi_{t-1,j}, \quad k=1, \ldots, K-1; t=2,\ldots, n.
\end{equation*}
Rewrite \eqref{eqDep}, the conditional probability $\pi_{t,k}^{t-1}$ can then be obtained by
\begin{equation}\label{eqCondProb}
\pi_{t,k}^{t-1}=\frac{\text{exp}(\delta_{t,k}+\sum_{j=1}^{K-1}\xi_{kj}Y_{t-1,j})}{1+\sum_{k=1}^{K-1}\text{exp}(\delta_{t,k}+\sum_{j=1}^{K-1}\xi_{kj}Y_{t-1,j})}, \quad k=1, \ldots, K-1; t=2,\ldots,n,
\end{equation}
and $\pi_{t,K}^{t-1}=\bigg(1+\sum_{k=1}^{K-1}\text{exp}[\delta_{t,k}+\sum_{j=1}^{K-1}\xi_{kj}Y_{t-1,j}]\bigg)^{-1}$.

In the MTM specified with \eqref{eqeta} and \eqref{eqDep}, we described the marginal mean regression in \eqref{eqeta} and the dependence model in \eqref{eqDep} separately to obtain a fully specified parametric model for correlated ordinal time series. This allows the interpretation of parameters in the marginal mean regression model in \eqref{eqeta} to be invariant to the specification of dependence structure. Moreover, the regression parameters in the marginal cumulative logit model in \eqref{eqeta} are orthogonal to the dependence parameters in the log local odds ratio model in \eqref{eqDep} followed by \cite{lee2007class}. The implication of orthogonality is that the maximum likelihood estimators of the regression parameters remain consistent even if the dependence model is incorrectly specified.

Another advantage of using the MTM is that the likelihood inference is available. We now derive the likelihood function.

Let $\bm{\theta}=(\alpha_{1}, \ldots, \alpha_{K-1}, B_{1}, \ldots, B_{K-1}, D_{1}, \ldots, D_{K-1}, \beta_{1}, \ldots, \beta_{K-1}, \Delta_{1}, \ldots, \Delta_{K-1})'$ be the $5(K-1)$ dimensional mean regression model parameter vector in \eqref{eqeta} for a given changepoint time $\tau$ and $\bm{\xi}=(\xi_{kj})_{k,j=1}^{K-1}$ be the $(K-1)^{2}$ dimensional dependence parameter vector. For the proposed MTM, the likelihood function consists of two components: the marginal probability of $Y_{t}$ at time $t=1$, $L_{1}(\bm{\theta})$, and the conditional probabilities of $Y_{t}$ given $Y_{t-1}$ for $t>1$, $L_{2}(\bm{\theta},\bm{\xi})$. Thus, the likelihood function up to constant can be obtained as
$L(\bm{\theta},\bm{\xi})=L_{1}(\bm{\theta})L_{2}(\bm{\theta},\bm{\xi})$, where 
\begin{equation*}\label{eqLogL1}
L_{1}(\bm{\theta})=P(Y_{1}=y_{1})=\prod_{k=1}^{K}\pi_{1,k}^{I_{(y_{1}=k)}}=\prod_{k=1}^{K}\pi_{1,k}^{y_{1,k}},
\end{equation*}
and
\begin{equation*}\label{eqLogL2}
L_{2}(\bm{\theta}, \bm{\xi})=\prod_{t=2}^{n}P(Y_{t}=y_{t}|Y_{t-1}=y_{t-1})=\prod_{t=2}^{n}\prod_{k=1}^{K}(\pi_{t,k}^{t-1})^{y_{t,k}},
\end{equation*}
where $y_{t}$ denote the observed data values. As before, $y_{t,k}=1$ if $y_{t}=k$ and $y_{t,k}=0$ otherwise, for $k=1, \ldots, K-1$. Note that $\sum_{k=1}^{K}y_{t,k}=1$. 

The maximum likelihood estimates of $\bm{\theta}$ and $\bm{\xi}$ are obtained by numerically maximizing the log likelihood function
\begin{align}\label{eqLogL}
\text{log} L(\bm{\theta}, \bm{\xi})=& \text{ log} L_{1}(\bm{\theta})+\text{log} L_{2}(\bm{\theta},\bm{\xi}) \nonumber \\
=&\bigg(y_{1,1}\text{log}(\gamma_{1,1})+\sum_{k=2}^{K-1}y_{1,k}\text{log}(\gamma_{1,k}-\gamma_{1,k-1})+y_{1,K}\text{log}(1-\gamma_{1,K-1})\bigg) \nonumber \\
& +\sum_{t=2}^{n}\sum_{k=1}^{K-1}y_{t,k}\text{log}(\pi_{t,k}^{t-1})
\end{align}
where $\gamma_{t,k}$ and $\pi_{t,k}^{t-1}$ can be found in \eqref{eqgammCalc} and \eqref{eqCondProb}, respectively.

\subsection{Parameter Estimation \label{ParameterEstimation}}
Maximizing the log-likelihood function is not a trivial computation job. In particular, the parameter vector $\bm{\phi}=(\bm{\theta}, \bm{\xi})$ contains $5(K-1)+(K-1)^2$ parameters to be estimated. The Fisher-scoring algorithm has been used for parameter estimation in the MTM for longitudinal categorical data \citep{lee2007class, lee2010longitudinal, lee2019marginalized}. 
However, unlike in longitudinal data, computing the information matrix for the Fisher-scoring algorithm requires significantly more time for a longer time series. In addition, the computational complexity increases exponentially with the number of categories $K$, potentially leading to computational infeasibility. In this study, a new parameter estimation procedure is developed to maximize the log-likelihood function in \eqref{eqLogL}.

The aforementioned orthogonality between $\bm{\theta}$ and $\bm{\xi}$ enables us to estimate them separately if there is no missing data \citep{lee2007class, lee2010longitudinal}. Specifically, the maximum likelihood estimates can be obtained by a Quasi-Newton algorithm \citep{giudici2013wiley} with updating $\bm{\theta}$ and $\bm{\xi}$ separately in each iteration. The estimation procedure works as follows. 

First,  the initial values for $\bm{\theta}$, denoted by $\bm{\theta}^{(0)}$, are obtained by fitting the marginal mean model in \eqref{eqeta} using the method in \cite{lu2012extended}. Let $\bar{\delta}_{k}$ be the average of the fitted cumulative probabilities $\gamma_{t,k}$, which will be used to derive $\bm{\xi}^{(0)}$, the initial values for $\bm{\xi}$.  Here, the values of $\bm{\xi}^{(0)}$ are calculated using the dependence model in \eqref{eqDep} as
\begin{equation*}\label{eqinitialxi}
    \xi_{kj}^{(0)} = \text{log}\bigg(\frac{\hat{p}_{jk}}{\hat{p}_{jK}}\bigg) - \bar{\delta}_{k}, \quad j=1,\ldots,K;k=1,\ldots,K-1,
\end{equation*}
where the transition probability $\hat{p}_{jk}$ is estimated as the ratio 
\begin{equation*}
    \hat{p}_{jk}=\frac{\{\# \ t:Y_{t-1}=j \cap Y_{t}=k \}}{\{\# \ t: Y_{t-1}=j\}}.
\end{equation*}
Observe that $\hat{p}_{jk}$ is simply the empirical proportion of times that category $j$ observed at time $t-1$ is followed by category $k$ at time $t$. It is worth noting that good initial values typically lead to faster convergence, requiring fewer iterations.

Next, we iteratively update $\bm{\theta}$ and $\bm{\xi}$ using the initial values $\bm{\theta}^{(0)}$ and $\bm{\xi}^{(0)}$. Convergence of the joint estimates is declared when
\begin{equation}\label{phiconverg}
\sqrt{(\hat{\bm{\phi}}^{\text{new}} - \hat{\bm{\phi}}^{\text{old}})^{'}(\hat{\bm{\phi}}^{\text{new}} - \hat{\bm{\phi}}^{\text{old}})} \leq \epsilon,
\end{equation}
where $\epsilon = 10^{-5}$ is the tolerance level for a relatively strict convergence criterion. Here, $\hat{\bm{\phi}}^{\text{old}}$ and $\hat{\bm{\phi}}^{\text{new}}$ denote the parameter estimates before and after the current iteration, respectively. In practice, convergence is typically achieved within one or two iterations. The detailed estimation procedure is provided in Section S1 of the supplementary materials. 

Once the parameter vector $\bm{\phi}$ has been estimated, denoted by $\hat{\bm{\phi}}$, its variance-covariance matrix can be approximated by the inverse of the Hessian matrix \citep{lee2016analysis}. Specifically, the Hessian matrix is estimated as
\begin{equation}\label{modelHessian}
I_{\bm{\phi}}(\hat{\bm{\phi}}; \bm{Y}) = \sum_{t=1}^{n} \left( \frac{\partial \log L(\hat{\bm{\phi}}; y_t)}{\partial \bm{\phi}} \right) \left( \frac{\partial \log L(\hat{\bm{\phi}}; y_t)}{\partial \bm{\phi}} \right)^{'},
\end{equation}
where the first-order derivatives of the log-likelihood function are given by
\begin{equation*}
\frac{\partial \log L(\bm{\phi}; \bm{Y})}{\partial \bm{\phi}} = \frac{\partial \log L_1(\bm{\phi}; \bm{Y})}{\partial \bm{\phi}} + \frac{\partial \log L_2(\bm{\phi}; \bm{Y})}{\partial \bm{\phi}}.
\end{equation*}

The explicit forms of these derivatives are provided in Section S2 of the supplementary materials. In the estimation procedure described above, the parameters ${\delta_{t,k}}$ are iteratively estimated using the Newton–Raphson algorithm, as detailed in Section S3. Moreover, in each iteration of updating either $\bm{\theta}$ or $\bm{\xi}$ (see Steps 1 and 2 in Section S1), all values of ${\delta_{t,k}}$ must be recalculated and hence, it requires significant computing time. Nevertheless, the proposed estimation method remains computationally efficient compared to approaches that update $\bm{\theta}$ and $\bm{\xi}$ simultaneously.

To assess the performance of our proposed estimation procedure, we conducted a simulation study in the absence of changepoints. 50,000 categorical time series of length $n=3650$ were generated with $K=5$ categories and the period $T=365$. Given the true values of $\bm{\theta}$ tabulated in Table \ref{tab1}, all categories have equal expected frequencies when the marginal categorical probabilities are constant, and the dependence structure exhibits a moderate positive correlation. 

In each simulation run, $y_1$ is generated from a multinomial distribution with categorical probabilities  $\pi_{1,1}, \ldots, \pi_{1,K}$.  Given $Y_{1}=y_1$, $y_2$ is simulated from a multinomial distribution with conditional categorical probabilities $\pi_{2,1}^{2-1},\ldots, \pi_{2,K}^{2-1}$.  Similarly, for $t=2, \ldots, n$, $y_t$ is drawn from a multinomial distribution with conditional probabilities $(\pi_{t,1}^{t-1},\ldots, \pi_{t,K}^{t-1})'$, where $\pi_{t,k}^{t-1}=P(Y_{t}=k|Y_{t-1}=y_{t-1})$. In all simulations in this study, the number of trials for the multinomial distribution is set to 1.

\begin{table}[h!]
\caption{Parameter estimates for $K=5$, $n=3650$ and $T=365$}\label{tab1}
\centering
\begin{tabular*}{\textwidth}{@{\extracolsep{\fill}}crrrr}
\toprule
\multicolumn{5}{c}{Marginal Mean Model} \\
        \midrule
        {\multirow{2}{*}{Parameter}} 
            & \multicolumn{1}{c}{\multirow{2}{*}{True}} 
                & \multicolumn{1}{c}{\multirow{2}{*}{Mean}} 
                    & \multicolumn{1}{c}{\multirow{1}{*}{Relative}} 
                        & \multicolumn{1}{c}{\multirow{1}{*}{Standard}} \\
         &   &   & \multicolumn{1}{c}{\multirow{1}{*}{Bias}}  &  \multicolumn{1}{c}{\multirow{1}{*}{Deviation}}\\
        \midrule
        {$\alpha_{1}$}     & -1.3863 & -1.3806 & 0.0041 & 0.1023 \\
        {$\alpha_{2}$}     & -0.4055 & -0.4008 & 0.0116 & 0.0851 \\
        {$\alpha_{3}$}     & 0.4055 & 0.4106 & -0.0126 & 0.0842 \\
        {$\alpha_{4}$}     & 1.3863 & 1.3928 & -0.0047 & 0.1004 \\
        {$B_{1}$}     & -0.1000 & -0.0955 & 0.0449 & 0.0715 \\
        {$B_{2}$}     & -0.2000 & -0.1975 & 0.0123 & 0.0598 \\
        {$B_{3}$}     & -0.1500 & -0.1479 & 0.0142 & 0.0593 \\
        {$B_{4}$}     & -0.3000 & -0.2986 & 0.0046 & 0.0716 \\
        {$D_{1}$}     & 0.2000 & 0.2007 & -0.0035 & 0.0721 \\
        {$D_{2}$}     & 0.1000 & 0.1004 & -0.0042 & 0.0601 \\
        {$D_{3}$}     & 0.1500 & 0.1504 & -0.0025 & 0.0599 \\
        {$D_{4}$}     & 0.3000 & 0.3009 & -0.0028 & 0.0718 \\
        {$\beta_{1}$}     & 0.1000 & 0.0868 & 0.1317 & 0.1758 \\
        {$\beta_{2}$}     & 0.1000 & 0.0919 & 0.0814 & 0.1470 \\
        {$\beta_{3}$}     & 0.1000 & 0.0931 & 0.0693 & 0.1461 \\
        {$\beta_{4}$}     & 0.1000 & 0.0946 & 0.0543 & 0.1745 \\
       \toprule
        \multicolumn{5}{c}{Dependence Model} \\
        \midrule
        {\multirow{2}{*}{Parameter}} 
            & \multicolumn{1}{c}{\multirow{2}{*}{True}} 
                & \multicolumn{1}{c}{\multirow{2}{*}{Mean}} 
                    & \multicolumn{1}{c}{\multirow{1}{*}{Relative}} 
                        & \multicolumn{1}{c}{\multirow{1}{*}{Standard}} \\
         &   &   & \multicolumn{1}{c}{\multirow{1}{*}{Bias}}  &  \multicolumn{1}{c}{\multirow{1}{*}{Deviation}}\\
        \midrule
        {$\xi_{11}$}     & 2.8000 & 2.8016 & -0.0006 & 0.2077\\
        {$\xi_{12}$}     & 2.2000 & 2.2051 & -0.0023 & 0.2077\\
        {$\xi_{13}$}     & 1.9000 & 1.9045 & -0.0024 & 0.2044\\
        {$\xi_{14}$}     & 1.0000 & 1.0005 & -0.0005 & 0.2182\\
        {$\xi_{21}$}     & 1.3000 & 1.2970 & 0.0023 & 0.1771 \\
        {$\xi_{22}$}     & 1.2000 & 1.1911 & 0.0074 & 0.1698 \\
        {$\xi_{23}$}     & 0.9000 & 0.8956 & 0.0049 & 0.1651\\
        {$\xi_{24}$}     & 0.6000 & 0.5952 & 0.0081 & 0.1665\\
        {$\xi_{31}$}     & 0.8000 & 0.7958 & 0.0053 & 0.1777\\
        {$\xi_{32}$}     & 0.8000 & 0.7956 & 0.0055 & 0.1675\\
        {$\xi_{33}$}     & 0.5000 & 0.4896 & 0.0208 & 0.1639\\
        {$\xi_{34}$}     & 0.7000 & 0.6973 & 0.0038 & 0.1550\\
        {$\xi_{41}$}     & 0.5000 & 0.4951 & 0.0098 & 0.1768\\
        {$\xi_{42}$}     & 0.4000 & 0.3932 & 0.0170 & 0.1685\\
        {$\xi_{43}$}     & 0.3000 & 0.2938 & 0.0207 & 0.1582\\
        {$\xi_{44}$}     & 0.3000 & 0.2881 & 0.0397 & 0.1540\\
\bottomrule
\end{tabular*}
\end{table}

The simulation results are summarized in Table~\ref{tab1}. The empirical means of the estimated parameters closely align with their true values, and the corresponding relative biases are minimal. Among the marginal mean parameters, the trend parameter estimators exhibit greater variability compared to the others. Additionally, the empirical standard deviations of $\hat{\bm{\xi}}$ are, on average, slightly larger than those of $\hat{\bm{\theta}}$.

Figures~S2 and S3 in the supplementary materials display histograms of $\hat{\bm{\theta}}$ and $\hat{\bm{\xi}}$, respectively, with the true parameter values indicated by red vertical lines. These histograms exhibit symmetry and bell-shaped curves, suggesting that the estimators are approximately normally distributed. This observation is further supported by the QQ plots shown in Figures~S4 and S5, which confirm the approximate normality of both $\hat{\bm{\theta}}$ and $\hat{\bm{\xi}}$.

\section{The $\Lambda_{\mbox{max}}$ Test}\label{LambdaMax}
This section introduces the maximally selected likelihood ratio changepoint test for correlated categorical time series. 
% This test is similar to the test used in \cite{lu2012extended}. 

\subsection{Likelihood Ratio Changepoint Test}\label{LRT}
Detecting a single changepoint in the marginal mean model in \eqref{eqeta} is equivalent to testing the null hypothesis
\begin{flalign*}
    & \hspace{2cm} \text{H}_{0}: \Delta_{k}=0 \text{ for all } k \in \{1, \ldots, K-1\}, &
\end{flalign*}
against the alternative hypothesis
\begin{flalign*}
    & \hspace{2cm} \text{H}_{a}: \Delta_{k}\neq 0 \text{ for some or all } k \in \{1, \ldots, K-1\}. &
\end{flalign*}

Let $\hat{\bm{\phi}}^{(0)}=(\hat{\bm{\theta}}^{(0)}, \hat{\bm{\xi}}^{(0)})$ and $\hat{\bm{\phi}}^{(a)}=(\hat{\bm{\theta}}^{(a)}(\tau), \hat{\bm{\xi}}^{(a)}(\tau))$ denote the maximum likelihood estimators of $\bm{\phi}$ for the null model and alternative model with a changepoint at time $\tau$, respectively. The log-likelihood ratio statistic is defined as 
\begin{equation*}
	\Lambda(\tau)= -2\bigg(\text{log} L(\hat{\bm{\phi}}^{(0)})-\text{log} L(\hat{\bm{\phi}}^{(a)}(\tau))\bigg).
\end{equation*}
This statistic is large when the null model fits poorly compared to the alternative model. \cite{wilks1935likelihood, wilks1938large} showed that $\Lambda(\tau)$ follows a Chi-square limiting distribution under $\text{H}_{0}$ for a known changepoint time $\tau$. However, with an unknown changepoint time, these statistics no longer have a Chi-square limiting distribution \citep{betensky1999maximally}. To detect the unknown changepoint,  we define the test statistic as \begin{equation}\label{eqLambdaMax}
    \Lambda_{\text{max}}=\max_{\ell\leq\tau/n\leq h}\Lambda(\tau),
\end{equation}
where $\Lambda(\tau)$ is maximized over a truncated set of admissible changepoint locations with $0 <\ell < h <1$.  The values of $\ell$ and $h$ are typically chosen to be close to 0 and 1, respectively,  in order to make most changepoint times admissible. In practice, detecting a changepoint near the beginning or end of the series is often unreliable due to the limited number of observations available before or after the changepoint.  This issue is especially pronounced for our MTM,  which involves estimating a relatively large number of parameters. Additionally, under the null hypothesis of no changepoints, the statistic $\max_{1 \leq  \tau  < n}\Lambda(\tau)$ converges to infinity as $n \rightarrow \infty$ when the maximization is taken over the full set $\{1,\ldots, n-1\}$.  

With boundary times cropped, for independent categorical data with $K$ categories, \cite{robbins2011changepoints} show that the maximally selected Pearson $\chi^2$ test statistic converges in distribution to 
\begin{equation}\label{supBB}
    \sup_{\ell \leq t \leq h} \frac{B^{(K-1)}(t)}{t(1-t)}.
\end{equation}
Here, $B^{K-1}(t)=\sum_{i=1}^{K-1}B^2_i(t)$, and $\{B_1(t), 0 \leq t\leq 1\}, \ldots, \{B_{K-1}(t), 0 \leq t \leq 1\}$ are independent Brownian bridge processes. Note that when linear trends are absent from the mean function in \eqref{eqeta}, the $\chi^2$ statistic is asymptotically equivalent to the likelihood ratio test statistic $\Lambda(\tau)$ for a known changepoint $\tau$ \citep   {agresti2002categorical}. As a result, the $\Lambda_{\text{max}}$ test for dependent categorical time series without trends is expected to have the same asymptotic distribution as in \eqref{supBB}. This expectation is based on the fact that the asymptotic theory for detecting mean shifts in independent data extends to weakly dependent time series \citep{csorgo1997limit}, which holds here in the MTM framework. A formal proof of this asymptotic equivalence is considerably more involved and remains an open problem requiring further theoretical investigation. Nevertheless, this theoretical result is supported by the simulated percentiles of the $\Lambda_{\text{max}}$ test presented below. More accurate percentiles are generally obtained through simulation.  

The empirical percentiles of the $\Lambda_{\text{max}}$ test in \eqref{eqLambdaMax} are reported in Table \ref{EmPercentile} for two marginal mean models with $K=5$ categories. The first model includes a seasonal component,  specified as $\eta_{t,k}=\alpha_{k}+s_{t,k}$. The second model, considered the full model, includes both a seasonal component and a scaled linear trend: $\eta_{t,k}=\alpha_{k}+s_{t,k}+\beta_{k}\frac{t}{n}$. 
Each entry in Table \ref{EmPercentile} is based on 10,000 independent simulations of $\Lambda_{\text{max}}$, with sample size $n=NT$, where $T=365$ and $N=3,5,10,20,30$ denotes the number of years. 
The parameter values used to generate this table are taken from Table \ref{tab1}. Since the empirical percentiles of $\Lambda_{\text{max}}$ are generally unaffected by changes in the period $T$, we fix $T=365$ for all simulations in this study. Additionally, we set $\ell= 0.05  $ and $h=  0.95 $,  as is common in the literature.

\begin{table}[h]
\caption{The empirical percentiles of $\Lambda_{\text{max}}$ for $K=5$ and various sample sizes $n=NT$ with $T=365$. Values were aggregated from 10,000 independent simulations}
\label{EmPercentile}
	\centering
	\begin{tabular*}{\textwidth}{@{\extracolsep{\fill}}ccccccc}
			\toprule
			\multirow{2}{*}{Marginal mean Model} & \multirow{2}{*}{Percentile} & \multicolumn{5}{c}{$n$}  \\ \cmidrule{3-7} 
			& & \multicolumn{1}{c}{$3T$} & \multicolumn{1}{c}{$5T$} & \multicolumn{1}{c}{$10T$} & \multicolumn{1}{c}{$20T$} & \multicolumn{1}{c}{$30T$} \\
			\midrule 
			\multirow{3}{*}{$\eta_{t,k}=\alpha_{k}+s_{t,k}$} & 90 & 15.743 & 15.903 & 15.838 & 15.825 & 15.840  \\
			& 95 & 17.918 & 17.911 & 17.808 & 17.764 & 17.757\\
			& 99 & 22.273 & 22.065 & 22.078 & 22.003 & 22.086\\
			\midrule 
			\multirow{3}{*}{$\eta_{t,k}=\alpha_{k}+s_{t,k}+\beta_{k}\frac{t}{n}$} & 90 & 17.903 & 17.767 & 17.792 & 17.755 & 17.666\\ 
			& 95 & 19.836 & 19.599 & 19.798 & 19.794 & 19.453 \\ 
			& 99 & 24.283 & 23.821 & 24.316 & 23.504 & 23.870 \\
			\bottomrule
		\end{tabular*}
\end{table}

For the marginal mean model with seasonal effects only, the reported empirical percentiles of $\Lambda_{\text{max}}$ closely match the corresponding 90th, 95th, and 99th percentiles of the asymptotic distribution in \eqref{supBB} from \cite{robbins2011changepoints}, which are 16.005, 17.892, and 22.026, respectively, under the null hypothesis of no changepoints. These percentiles may also be used for stationary series with $\eta_{t,k}=\alpha_k$. When linear trends are added to the marginal mean model, the empirical percentiles become larger than those for the marginal mean model without trends at each nominal level. As with CUSUM tests, ignoring trends will deflate the asymptotic percentiles \citep{gallagher2013changepoint, robbins2016general}. Importantly, additional simulation studies show that the empirical percentiles of $\Lambda_{\text{max}}$ are invariant to different choices of parameter vector $\bm{\phi}$, provided that the number of categories $K$ is fixed and the truncated set $\ell\leq\tau/n\leq h$ remains unchanged.

Finally, the null hypothesis $\text{H}_{0}$ is rejected when the $\Lambda_{\text{max}}$ value exceeds the $100(1-\alpha)\%$ percentile at significance level $\alpha$. The maximum likelihood estimator of the changepoint time, $\hat{\tau}$, is the value of $\tau$ that maximizes $\Lambda(\tau)$ over $\ell\leq\tau/n\leq h$.

%{\color{red} Lastly, the significance of $\Lambda_{\text{max}}$ may arise from contributions of either all or a subset of categories, depending on the underlying pattern of distribution changes. Some categories may exhibit significant changes in marginal probabilities, while others remain stationary and contribute less to the $\Lambda_{\text{max}}$ test statistic. To dissect these dynamics and examine the specific contributions of each category, the significance of individual $\Delta_{k}$ parameters associated with the estimated changepoint $\hat{\tau}$ can be assessed using the Wald inference described in supplementary materials Section~S3.}

\section{Simulation study}\label{simulationstudy} In this section, we present extensive simulation studies to explore the performance of our proposed method.

\subsection{Effects of Autocorrelation}\label{Acf}
Simulations in this section assess how autocorrelations impact the performance of the changepoint test. Basically, we compare the $\Lambda_{\text{max}}$ in Section \ref{LRT} for correlated categorical series and $\lambda_{\text{max}}$ in \cite{lu2012extended} for independent categorical data.

In each simulation, a series is generated with $n=1825$ and $K=5$. To study the effects of autocorrelation on Type I error rates only, we set the marginal mean model without seasonality and changepoints, that is $\eta_{t,k}=\alpha_{k}+\beta_{k}\frac{t}{n}$. The $\{\alpha_{k}\}$ and $\{\beta_{k}\}$ parameter values used for the simulation are those listed in Table~\ref{tab1}. Here, we consider five different sets of dependence parameters: $\bm{\xi}_{1}$, $\bm{\xi}_{2}$, $\bm{\xi}_{3}$, $\bm{\xi}_{4}$, and $\bm{\xi}_{5}$, which are listed in the footnotes of Table \ref{AcfEffect}. As the magnitude of each entry in $\bm{\xi}$ increases, the degree of serial correlation increases. The dependence parameters in $\bm{\xi}_{1}$ are all positive with the largest magnitudes, resulting in a simulated categorical time series with the strongest positive autocorrelation. Similarly, $\bm{\xi}_{5}$ will produce sequences with the strongest negative autocorrelation. $\bm{\xi}_{2}$ and $\bm{\xi}_{4}$ are associated with moderate positive and negative autocorrelations, respectively. When all parameters in $\bm{\xi}_{3}$ are equal to zero, the simulated series is independent. 

The empirical proportions of the 5000 simulated series that falsely reject the null hypothesis of no changepoint are reported in Table~\ref{AcfEffect}, at a nominal level of 5\% for all five different sets of $\bm{\xi}$. Both $\Lambda_{\max}$ and $\lambda_{\max}$ were applied to each simulated series. The $\Lambda_{\max}$ test declares a changepoint if the $\Lambda_{\max}$ statistic exceeds 19.6 from Table~\ref{EmPercentile} for $N=5$; the $\lambda_{\max}$ test rejects the null hypothesis of no changepoint if the $\lambda_{\max}$ statistic exceeds of 19.4, which is the 95th percentile of $\lambda_{\max}$ that we obtained by repeating the simulation in \cite{lu2012extended} with $K=5$ and $  0.05   \leq \tau/n \leq   0.95  $. The dataset features only one event occurring at each time $t$, whereby one category records a frequency of one, while all other categories register a frequency of zero.

\begin{table*}[!htbp]
\caption{Effects of autocorrelation on changepoint detection}\label{AcfEffect}
\centering 
\begin{threeparttable}
\centering 
\begin{tabular*}{\textwidth}{@{\extracolsep{\fill}}ccc}
\toprule Dependence & Type I error rates & Type I error rates\\
parameter & of $\Lambda_{\text{max}}$ & of $\lambda_{\text{max}}$\\
\midrule 
$\bm{\xi}_{1}$   & 0.062 & 0.670 \\
$\bm{\xi}_{2}$   & 0.051 & 0.316 \\
$\bm{\xi}_{3}$   & 0.050 & 0.050 \\
$\bm{\xi}_{4}$   & 0.048 & 0.040 \\
$\bm{\xi}_{5}$   & 0.046 & 0.042 \\
\bottomrule 
\end{tabular*}\par 
\begin{tablenotes}
\tiny
\item{$\bm{\xi}_{1}=\{5.9,  4.0,  2.6,  1.7,  4.7,  3.6,  2.4,  1.7,  3.2,  2.2,  1.5,  1.2,  1.8,  1.6,  0.9,  0.6\}$;}
\item{$\bm{\xi}_{2}=\{3.9,  2.7,  1.7,  1.3,  2.3,  1.5,  1.0,  0.5,  2.1,  1.1,  0.9,  0.7,  1.2,  0.9,  0.6,  0.5\}$;}
\item{$\bm{\xi}_{3}=\{0, 0, 0, 0, 0, 0, 0, 0, 0, 0, 0, 0, 0, 0, 0, 0\}$;}
\item{$\bm{\xi}_{4}=\{-3.9,  -2.7,  -1.7,  -1.3,  -2.3,  -1.5,  -1.0,  -0.5,  -2.1,  -1.1,  -0.9,  -0.7,  -1.2,  -0.9,  -0.6,  -0.5\}$;}
\item{$\bm{\xi}_{5}=\{-5.9,  -4.0,  -2.6,  -1.7,  -4.7,  -3.6,  -2.4,  -1.7,  -3.2,  -2.2,  -1.5,  -1.2,  -1.8,  -1.6,  -0.9,  -0.6\}$;}
\end{tablenotes}
\end{threeparttable}
\end{table*}

In Table~\ref{AcfEffect}, the Type I error rates of the $\Lambda_{\text{max}}$ test that takes into account autocorrelations are very close to the nominal 0.05 level, suggesting a well-performing test. In contrast,  the $\lambda_{\text{max}}$ test that ignores autocorrelations has much higher Type I error rates than 0.05 for positively correlated series ($\bm{\xi}_{1}$ and $\bm{\xi}_{2}$) as expected. When the positive autocorrelation gets stronger, the transition probabilities from one category to itself or nearby categories between adjacent time points, $p_{t,kk}=\text{P}(Y_{t}=k|Y_{t-1}=k)$, become larger and consequently the observed categories tend to hang together, which might be mistaken as a level shift.
Even for a moderate positive autocorrelation ($\bm{\xi}_{2}$), the Type I error rates of $\lambda_{\text{max}}$ is still about 6 times larger than $\Lambda_{\text{max}}$. Note that when there is no autocorrelation, the two tests perform equivalently. For negatively correlated series ($\bm{\xi}_{4}$ and $\bm{\xi}_{5}$), the $\lambda_{\text{max}}$ test is slightly more conservative than $\Lambda_{\text{max}}$. In short, ignoring positive autocorrelations can result in a higher probability of falsely declaring a changepoint, while ignoring negative autocorrelations may slightly reduce the power of detecting a true changepoint. A similar pattern was described in \cite{lund2007changepoint}.

\subsection{Detection power \label{power}}
The detection power of the changepoint test is evaluated through two simulation studies in this section. Specifically, we examine how often the methods successfully detect a changepoint when one is present. The parameter values ${\alpha_k}$, ${\beta_k}$, ${B_k}$, and ${D_k}$ used in the simulation are taken from Table~\ref{tab1}. 

In the first simulation study, we compare the detection power of our $\Lambda_{\text{max}}$ test with the $\lambda_{\text{max}}$ test from \cite{lu2012extended}. %Since the $\lambda_{\text{max}}$ test requires an independence assumption and data aggregation helps to reduce seasonality and serial correlation, we applied the $\lambda_{\text{max}}$ test to the data aggregated annually while applying our $\Lambda_{\text{max}}$ test to the simulated daily data.
We generate $N=20$ years of daily correlated categorical time series with $K=5$ categories, introducing a single changepoint. The total length of the time series is  $n=365N=7300$.  We consider the dependence parameter sets $\bm{\xi}_{1}$, $\bm{\xi}_{2}$, and $\bm{\xi}_{3}$, which correspond to strongly positively autocorrelated, moderately positively autocorrelated, and independent scenarios, as negative autocorrelations are rarely observed in climate time series.  The changepoint time $\tau$ is introduced at a randomly selected location within the range $\ell= 0.05 \leq\tau/n \leq h= 0.95 $. The magnitude of changes is the same across all categories: $\Delta_{1}=\ldots=\Delta_{K-1}=\Delta$, where $\Delta=0, 0.3, 0.5$. We then apply both the $\Lambda_{\text{max}}$ and $\lambda_{\text{max}}$ tests. Unlike in Section~\ref{Acf},  the simulated 20 years of daily data are aggregated to annual frequencies to compute the $\lambda_{\text{max}}$ test statistic, resulting in a total event count of 365 per year. Similar to the criterion used in \cite{lund2007changepoint},  a successful detection is counted if the estimated changepoint is within one year of the true changepoint for the $\lambda_{\text{max}}$ test and within 18 months (or 540 days) for the $\Lambda_{\text{max}}$ method when $|\Delta|>0$.

\begin{table}[!h]
\caption{Estimated detection powers of $\Lambda_{\text{max}}$ and $\lambda_{\text{max}}$ for $K=5$}\label{Power}
\centering
\begin{tabular*}{\textwidth}{@{\extracolsep{\fill}}lccccc}
\toprule
    Dependence Parameter & Test & $\Delta=0$ & $\Delta=0.3$ & $\Delta=0.5$ \\
    \midrule 
    \multirow{2}{*}{$\bm{\xi}_{1}$} & $\Lambda_{\text{max}}$ & 0.048 & 0.106 & 0.447 \\
    & $\lambda_{\text{max}}$ & 0.142 & 0.245 & 0.623 \\
    \midrule 
    \multirow{2}{*}{$\bm{\xi}_{2}$} & $\Lambda_{\text{max}}$ & 0.050 & 0.251 & 0.813 \\
    & $\lambda_{\text{max}}$ & 0.068 & 0.368 & 0.830 \\
    \midrule 
    \multirow{2}{*}{$\bm{\xi}_{3}$} & $\Lambda_{\text{max}}$ & 0.049 & 0.558 & 0.985 \\
    & $\lambda_{\text{max}}$ & 0.048 & 0.447 & 0.952 \\
\bottomrule
\end{tabular*}
\end{table}

Table \ref{Power} presents the empirical powers of both tests, collected from 1000 independent simulations at the nominal level of 5\%.  As expected, the detection power of both tests increases as $|\Delta|$ increases. Note that when $\Delta=0$, the Type I error rates are also reported in Table \ref{Power}. For independent data with $\bm{\xi}_{3}$, both tests exhibit Type I error rates close to the nominal 0.05 level. However, as the autocorrelation in the daily data increases from $\bm{\xi}_{2}$ to $\bm{\xi}_{1}$, aggregation fails to effectively reduce serial correlation, leading to an inflated Type I error rate for the $\lambda_{\text{max}}$ test. In contrast, our $\Lambda_{\text{max}}$ test maintains well-controlled Type I error rates. While annual aggregation helps reduce seasonality and autocorrelation, the independence assumption for $\lambda_{\text{max}}$ should be used with caution. When comparing the two tests for $\Delta=0.3$ and $0.5$, the $\lambda_{\text{max}}$ test appears slightly more powerful, possibly due to its higher Type I error rate. It is also known that tests with more parameters generally have lower power than those with fewer parameters.

In the second simulation study, we compare the detection power of our $\Lambda_{\text{max}}$ test with the CUSUM methods from \cite{li2022changepoint}, specifically the one-step-ahead prediction residuals method (denoted as $M_{I}$) and the reconstructed latent process-based method (denoted as $\tilde{M}_{\epsilon}$), demonstrating how our test performs better under various patterns of category changes.  

  $N=3$ years of daily correlated categorical time series were generated with $K=5$ categories and a single changepoint. The resulting sample size is $n=365N=1095$. Detection power is evaluated for both marginal mean models with and without trends under strong positive autocorrelation ($\bm{\xi}_{1}$). The true changepoint time, $\tau$, is specified at locations corresponding to proportions $\tau/n$, with values $\tau/n=0.2, 0.3, 0.4, 0.5, 0.6, 0.7$, and $0.8$.  Here, we consider six different $\Delta_{k}$ specifications: 
\begin{itemize}
    \item[] SC1: $\Delta_{1}=\Delta_{2}=\Delta_{3}=\Delta_{4}=\Delta$
    \item[] SC2:   $\Delta_{1}=\Delta_{2}=\Delta_{4}=\Delta$; $ \Delta_{3}=0$
    \item[] SC3: $\Delta_{1}=\Delta_{3}=0; \Delta_{2}=\Delta_{4}=\Delta$ 
    \item[] SC4: $\Delta_{1}=\Delta_{2}=0; \Delta_{3}=\Delta_{4}=\Delta$ 
    \item[] SC5: $\Delta_{1}=\Delta_{2}=\Delta_{3}=0; \Delta_{4}=\Delta$
    \item[] SC6: $\Delta_{1}=\Delta_{3}=\Delta; \Delta_{2}=\Delta_{4}= - \Delta$
\end{itemize}
with $\Delta = -0.3$, $0.3$, and $0.5$.
%The common changepoint parameter $\Delta$ is examined at three levels: $\Delta = -0.3$, $0.3$, and $0.5$. 
For each combination of $\tau$, $\Delta$, and the six $\Delta_k$ specifications, 1000 independent series were simulated. The detection power of the three tests ($\Lambda_{\text{max}}$, $M_{I}$, and $\tilde{M}_{\epsilon}$) is estimated as the empirical proportion of the 1000 simulated series that correctly identify a changepoint. The detection powers under the marginal model without trends are presented in Figure~\ref{comapreCUSUM}, while those under the marginal model with trends are shown in Figure S6 of the supplementary material.

\begin{figure*}[h!]
% \centerline{\includegraphics[scale=0.32]{figs/StrPosComparison_model1.eps}}
\centerline{\includegraphics[scale=0.31]{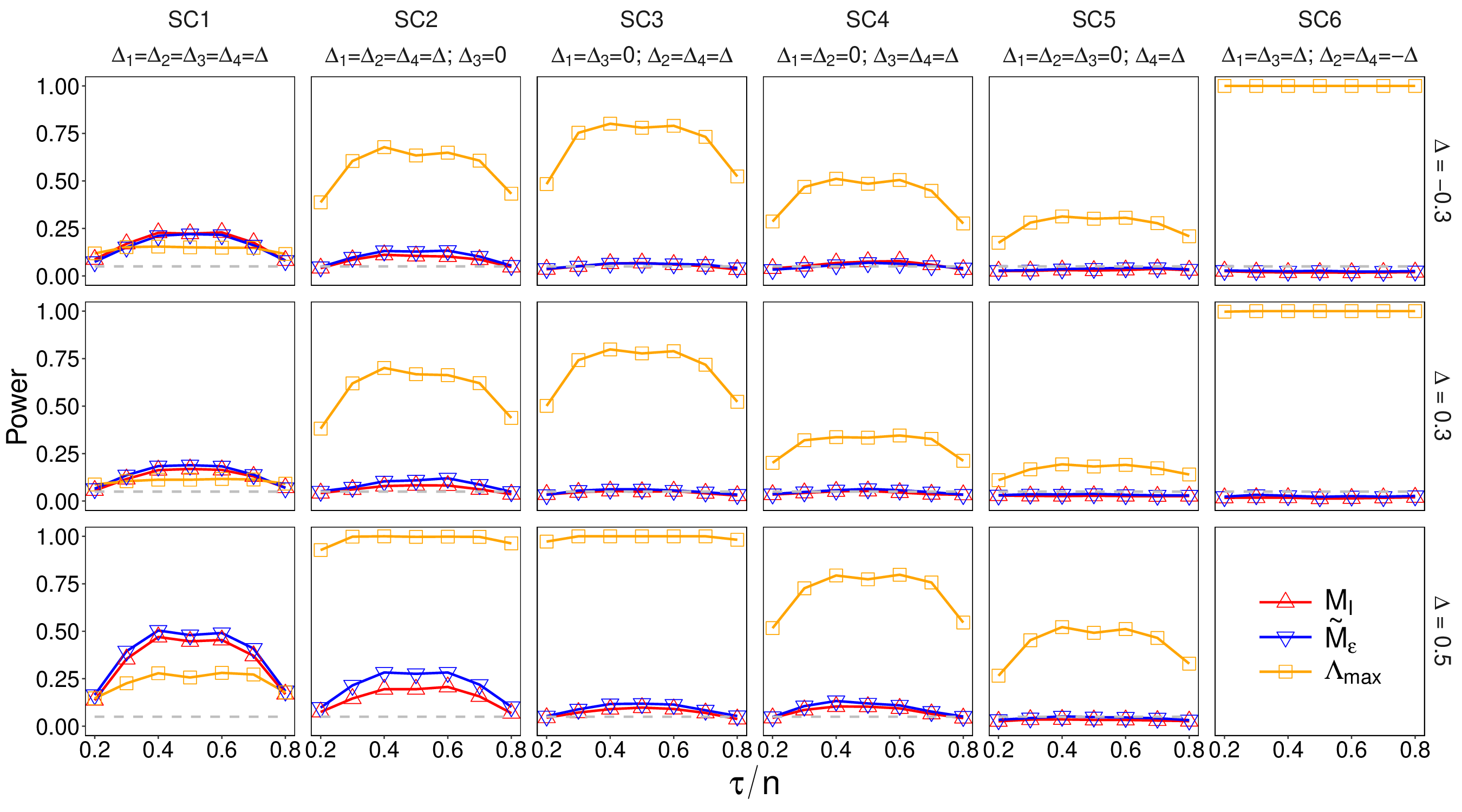}}
\caption{Estimated detection power of the $\Lambda_{\text{max}}$, $M_{I}$, and $\tilde{M}_{\epsilon}$ tests for the marginal model without trends under strong positive autocorrelation, plotted as a function of $\tau/n$ for three levels of $\Delta$. \label{comapreCUSUM}}
\end{figure*}

According to Equation~\eqref{marginalprob}, changes in the categorical probabilities for category $k=1$ are introduced via $\Delta_{1}$. For categories $2 \leq k \leq K-1$, changepoint effects on $\pi_{t,k}$ arise from both $\Delta_{k-1}$ and $\Delta_{k}$, reflecting the ordinal structure of the categories. Finally, $\Delta_{K-1}$ determines the change in $\pi_{t,K}$. In SC1, the changepoint affects the categorical probabilities $\pi_{t,k}$ across all five categories, with a common $\Delta$ applied to each category $k <K$. Specifically, SC1 shifts the categorical probabilities monotonically toward higher categories when $\Delta > 0$, and toward lower categories when $\Delta < 0$. This corresponds to the changepoint effect specification in \cite{li2022changepoint}. As shown in Figure~\ref{comapreCUSUM}, for all three levels of $\Delta$, the detection power of the two CUSUM tests is slightly higher than that of $\Lambda_{\text{max}}$ when the changepoint occurs in the middle of the time series. Near both ends of the series, all three tests perform similarly. However, in contrast to SC1, the CUSUM tests in \cite{li2022changepoint} have a limitation in detecting changepoints when the $\Delta_k$'s do not share a common level shift $\Delta$. This includes scenarios where changes in categorical probabilities occur only in a subset of categories, as in SC3, SC4, and SC5, or where all categories are affected but with alternating directions of change in adjacent categories, as in SC2 and SC6. Clearly, from Figure~\ref{comapreCUSUM}, Our $\Lambda_{\text{max}}$ test is substantially more powerful and practically superior to the CUSUM tests in \cite{li2022changepoint} under specifications SC2 to SC6, across all values of $\Delta$ and $\tau/n$.

Moreover, note that beyond the magnitude of the changes in categorical probabilities, the overall changepoint effect for our $\Lambda_{\text{max}}$ test also depends on whether adjacent $\Delta_{k}$ values have the same or opposite signs. For instance, in SC1, where adjacent $\Delta_{k}$ values share the same nonzero $\Delta$, the changepoint effect tends to diminish due to attenuation across adjacent categories, which degrades the test’s performance. In contrast, SC6 represents a more extreme scenario in which adjacent $\Delta_{k}$ values have the same magnitude but opposite signs. This configuration minimizes the attenuation of changepoint effects in the categorical probabilities of adjacent categories. As a result, the detection power of our $\Lambda_{\text{max}}$ test reaches its maximum value of 1 for $|\Delta| = 0.3$. The simulation results for $\Delta = 0.5$ under SC6 are not shown here, as they are nearly identical to those for $\Delta = 0.3$.

Finally, as expected, the detection power of all three tests increases with the magnitude of $|\Delta|$. In addition, the simulation results for the marginal mean model with trends, shown in Figure~S6, generally exhibit lower detection power compared to those for the marginal model without trends, which reduce the power of the tests. Furthermore, the detection power curves for all three tests under the mean model with trends partially exhibit an M-shaped pattern over the truncated ${\tau/n}$ set. An M-shaped detection power curve has also been observed in \cite{gallagher2013changepoint} and \cite{li2022changepoint} for models with linear trends.

\section{Total Cloud Cover at Fort St. John Airport}\label{Application}

The proposed method is applied to analyze two daily sky-cloudiness condition series recorded at 9 a.m. and 3 p.m. local standard time at Fort St. John Airport in Canada between 1965 and 1994. As noted in Section~\ref{data}, the associated metadata indicates a changepoint on November 7, 1979, due to a station relocation and a change in ownership. 

Figures~\ref{fig3} and S7 of the supplementary materials illustrate the daily categorical time series observed at 9 a.m. and 3 p.m., respectively, with five categories for the years 1978 and 1979. A seasonal cycle is visually evident, with more cloudy days than clear days during these two years. This is supported by the periodogram plots (not shown here), which display a prominent yearly period of $T=365$ for both daily series. To maintain a consistent duration of 365 days per year, leap days (February 29) were excluded from the series when they occurred. Each series comprises 
$n=30T=10,950$ data points, observed consistently at the same time (9 a.m. or 3 p.m.) each day.

\begin{figure*}[!h]
\centerline{\includegraphics[scale=0.35]{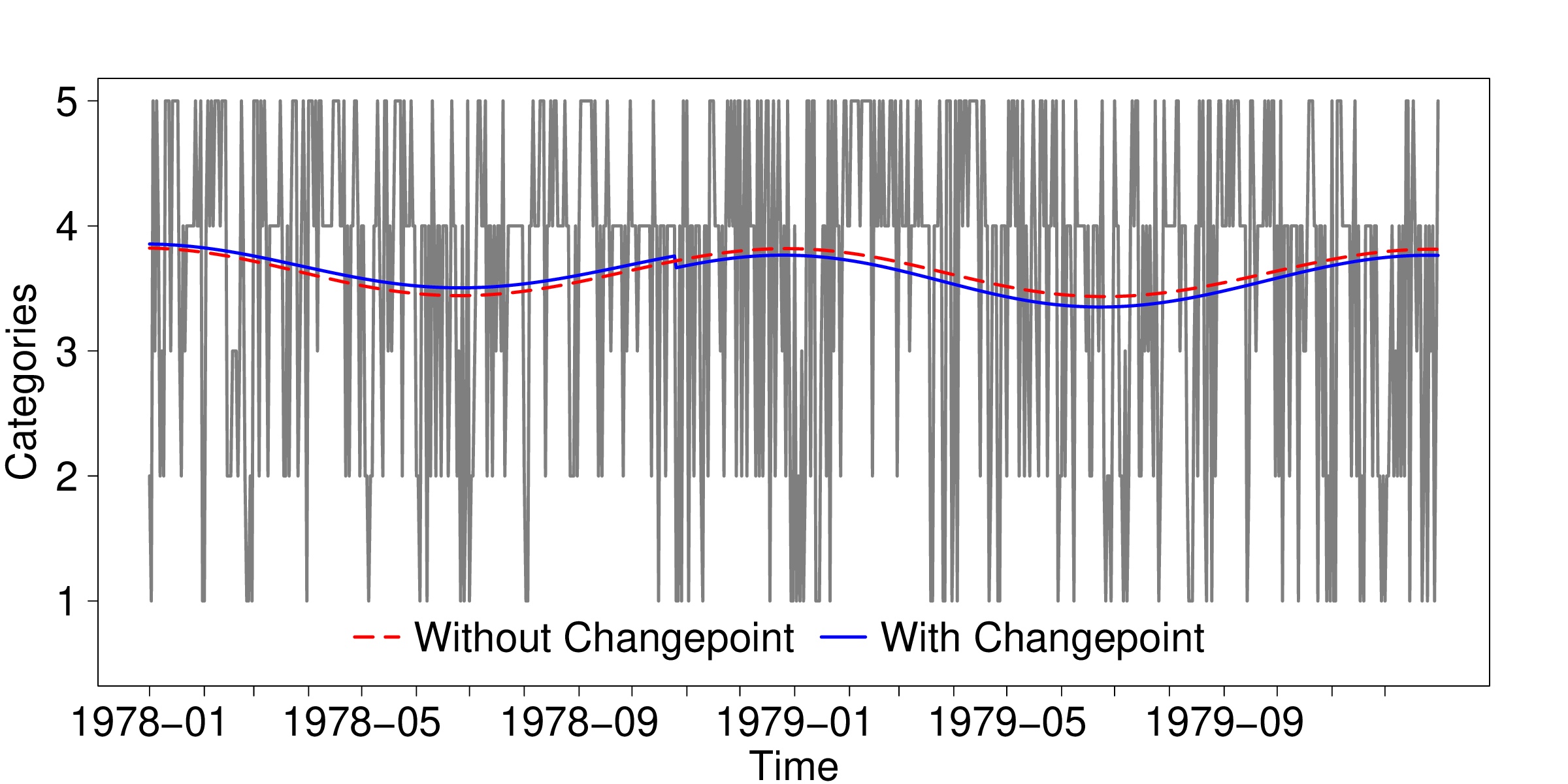}}
\caption{Daily sky-cloudiness conditions observed at 9 a.m. at Fort St. John Airport in 1978 and 1979, with the estimated mean and detected changepoint (October 26, 1978).\label{fig3}}
\end{figure*}

The $\Lambda_{\text{max}}$ test described in Section~\ref{LambdaMax} is applied to the two daily cloud cover time series using the marginal mean model in \eqref{eqeta}, which includes both a scaled linear trend and a seasonal component $s_{t,k}$. The seasonal component is approximated by a first-order Fourier expansion with a period of $T=365$, as specified in \eqref{eqalpha}. 

\subsection{Daily cloud cover series at 9 a.m. }
This subsection presents the results for the daily cloud cover series observed at 9 a.m.    

\begin{figure}[!h]
\centerline{\includegraphics[scale=0.29]{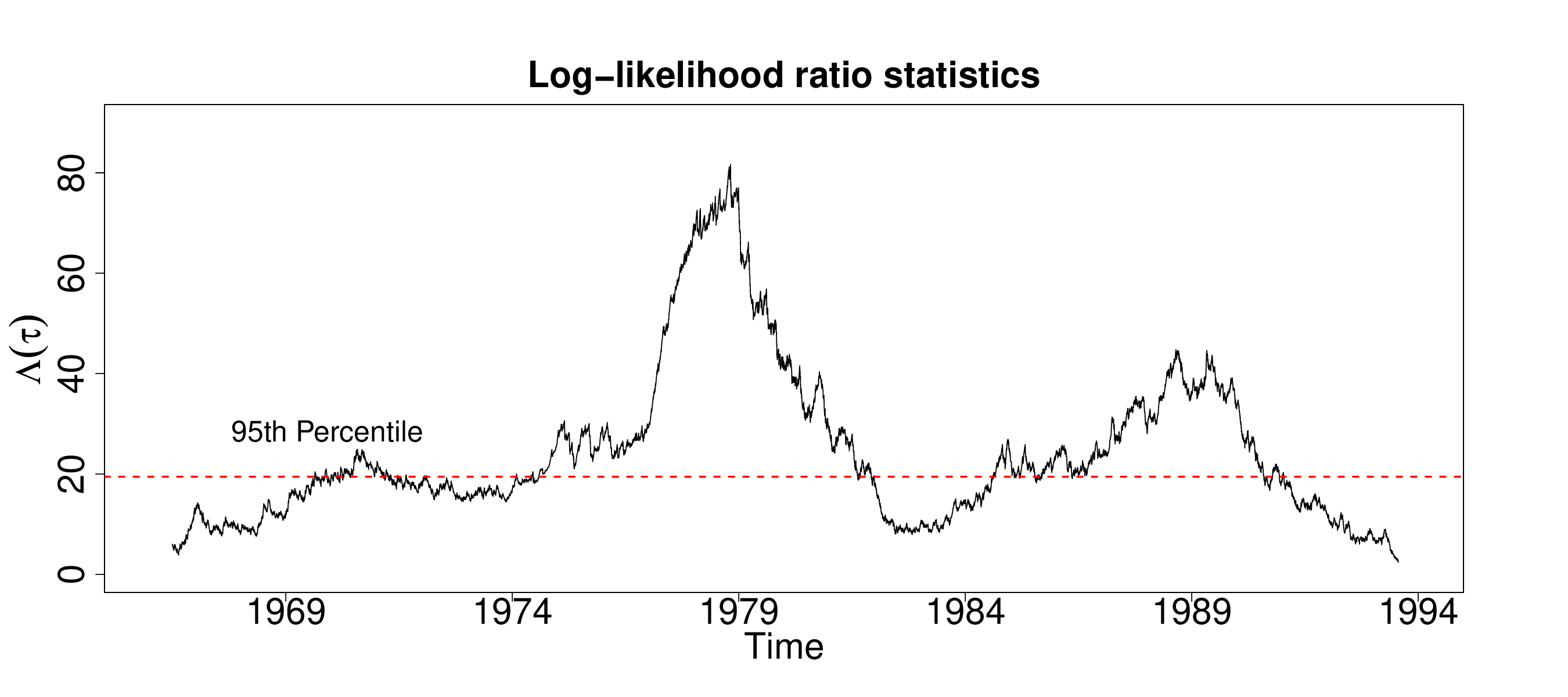}}
\caption{Log-likelihood ratio statistics $\Lambda(\tau)$  for daily sky-cloudiness conditions observed at 9 a.m. at Fort St. John Airport. \label{fig4}}
\end{figure}

 The log-likelihood ratio statistics $\Lambda(\tau)$ for each admissible changepoint $\tau$ in the cropped set $ 0.05 \leq\tau/n \leq   0.95  $, along with the 95th empirical percentile of $\Lambda_{\text{max}}$ (red dotted line), are plotted in Figure~\ref{fig4} for the daily series at 9 a.m. The largest value, $\Lambda_{\text{max}}=81.639$, occurs at $\hat{\tau}=5044$ (October 26, 1978), and greatly exceeds its 95th empirical percentile (19.453 in Table \ref{EmPercentile} for $N=30$), suggesting a significant change in the later days of 1978.

%We first applied the $\Lambda_{\text{max}}$ test described in Section~\ref{LambdaMax} by fitting the MTM model without changepoint specified in \eqref{eqeta}, consistent with Model 2 used in the simulations in Section~\ref{LRT}. This was applied to the five-category time series described earlier, including both a scaled linear trend and a seasonal component $s_{t,k}$. The seasonal component was approximated using a first-order Fourier expansion with a period of $T=365$, as specified in \eqref{eqalpha}.  

\begin{figure}[!h]
\centerline{\includegraphics[scale=0.31]{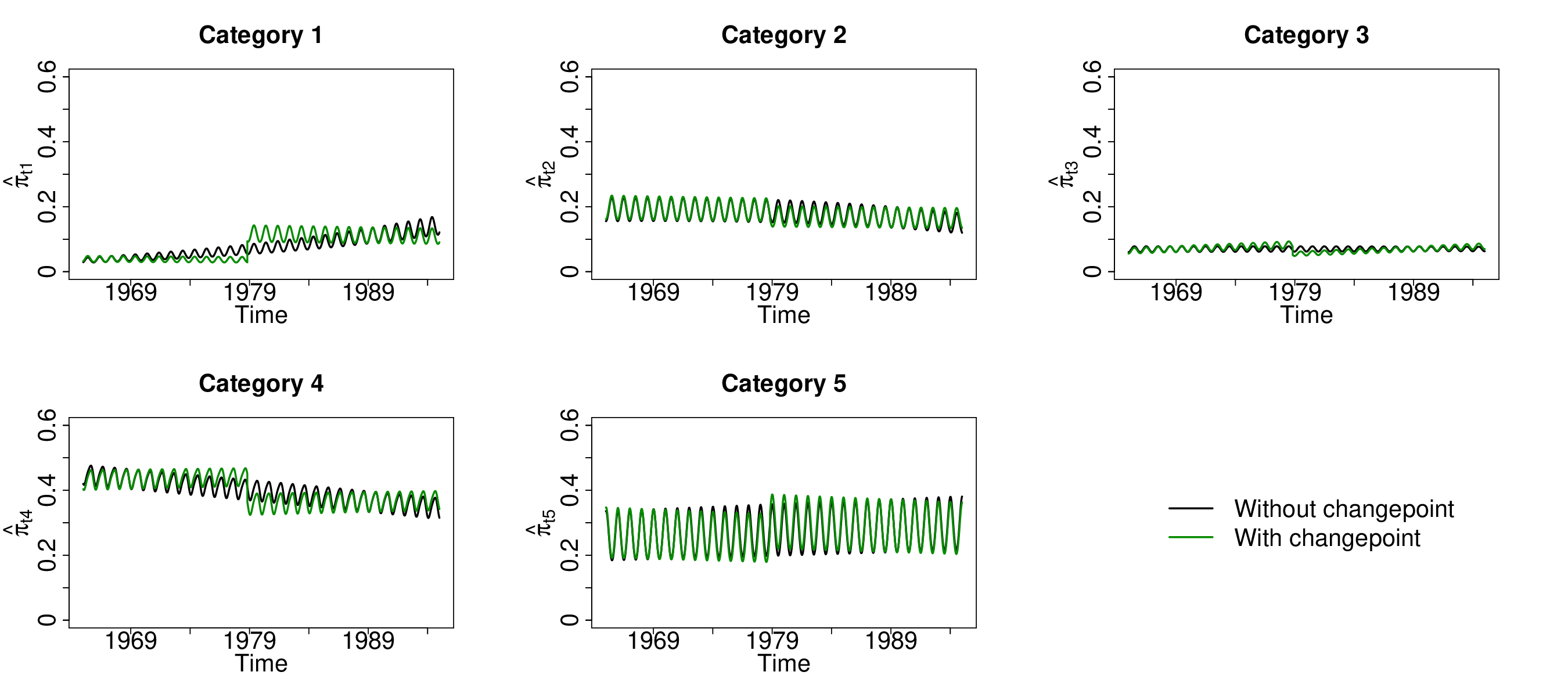}}
\caption{Estimated categorical probabilities $\hat{\pi}_{t,k}$ without a changepoint (black) and with a changepoint (green) across five categories for daily sky-cloudiness conditions observed at 9 a.m. at Fort St. John Airport. \label{fig7}}
\end{figure}

Additionally, the parameter estimates and their corresponding 95\% confidence intervals for the marginal mean models, with and without the estimated changepoint, are presented in the first two columns of Table S1 in the supplementary materials. The standard errors of the parameter estimates are derived from the evaluated Hessian matrix, as described in Section S4 of the supplementary materials. Note that the scaled linear trend parameters ($\beta_{k}$) are significant in the model without changepoints. However, in the model with the estimated changepoint at $\hat{\tau}=5044$, the changepoint parameters $\Delta_{k}$ are significant for $k=1,2,$ and $4$,  while the linear trend parameters are no longer significant. In practice, fitting a model without a changepoint when one is present generally leads to a masked structural change, appearing as a positive trend when the mean shifts are positive, or as a negative trend when the mean shifts are negative. This effect is illustrated in Figure~\ref{fig7}, which shows the estimated categorical probabilities $\hat{\pi}_{t,k}$ from the marginal mean models with and without the changepoint at $\hat{\tau}$ across all five categories. Moreover, consistent with the conclusion for the daytime cloud cover series in \cite{lu2012extended}, category 1 (clear sky) exhibited the most significant change due to the weather station relocation in 1979. Categories 2, 4, and 5 were also notably affected by the changepoint, while category 3 (scattered) was less impacted.

% 1. 
% y label into pi_tk (Done)
% legend without "Model 2" (Done)

% 2. 
% Figure S..... order needs to change (Done)

% 3.
% Table S1. change model 2 to "marginal mean model in equation (2)"; (Done)
%  drop Model 2 in row 2; (Done)

Furthermore, the fitted values of $Y_t$ for the years 1978 and 1979, obtained by $\hat{Y}_{t} = \sum_{k=1}^K k\hat{\pi}_{t,k}$, are plotted in Figure~\ref{fig3} for both marginal mean models—with and without the estimated changepoint. The average of absolute changes $|\hat{\Delta}_{k}|$ ($k = 1, \ldots, K-1$) is 0.494 in the logits of cumulative probabilities. In addition, as expected, the estimated dependence parameters $\bm{\hat{\xi}}$ are approximately the same for both models, regardless of whether the changepoint at $\hat{\tau} = 5044$ is included.
 
\begin{figure}[!h]
\centerline{\includegraphics[scale=0.29]{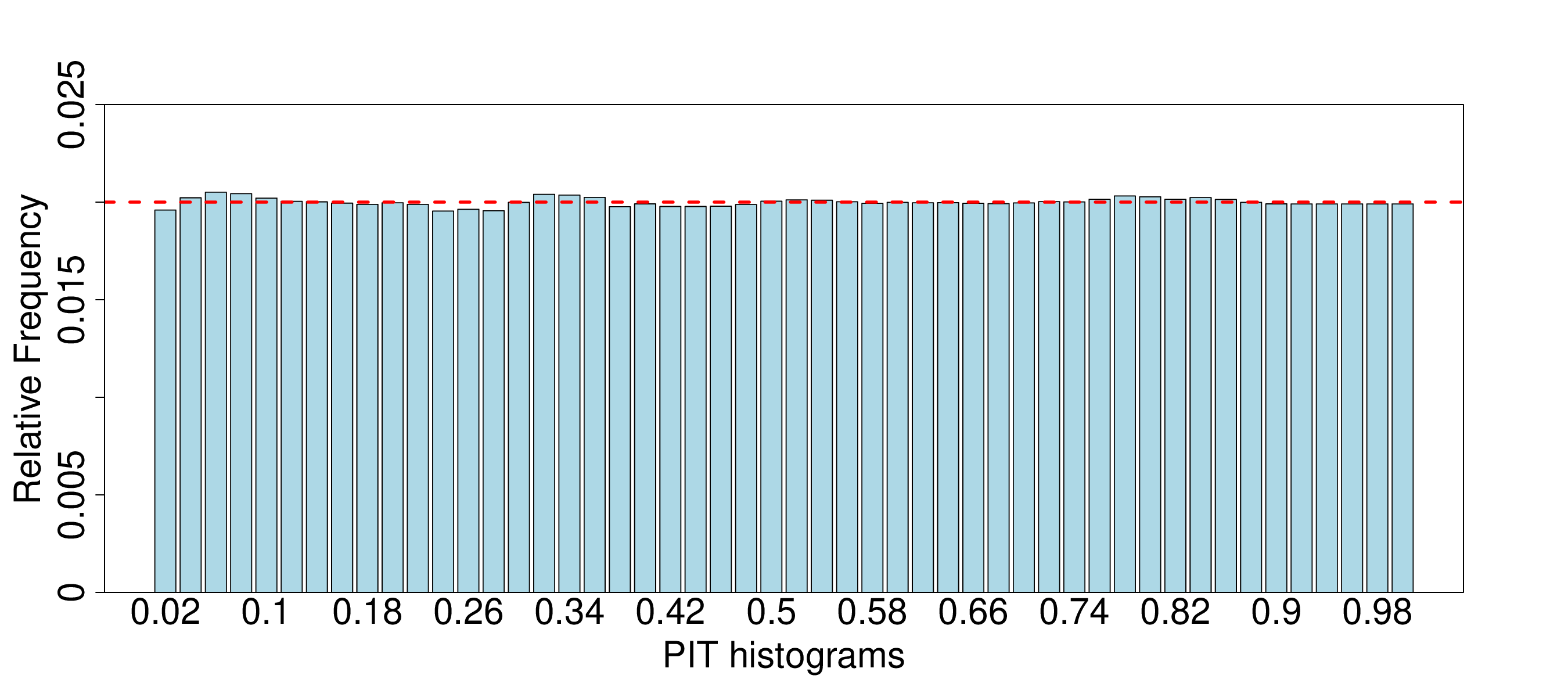}}
\caption{PIT histogram for the marginal mean model with the estimated changepoint for daily sky-cloudiness conditions observed at 9 a.m. at Fort St. John Airport. \label{fig5}}
\end{figure}

Lastly, to assess the adequacy of our model fit, probability integral transform (PIT) histograms, as described in Section~S5 of the supplementary materials, are employed. Given the large sample size of the Fort St. John Airport series, we use $H=50$ bins at points $h/50$, $h=1,\ldots,H$. The resulting PIT histogram for the 9 a.m. series, shown in Figure~\ref{fig5},  indicates that most bars align closely with the expected uniform probability of $1/H=0.02$ (horizontal dashed line in red), suggesting that the model is well calibrated.

%Due to the large sample size of the Fort St. John Airport series, we select $H=50$ bins at the points of $h/50$, $h=1,\ldots,H$. The constructed histogram implies that the model fits well by comparing the observed distribution of PIT values to the expected uniform distribution. The well-calibrated uniformity of the predictive distribution is indicated by the majority of bar heights closely aligning with the expected probability of $1/H=0.02$ (horizontal red dashed line) for the series observed at 9 a.m. in Figure~\ref{fig5}. This alignment suggests that the model's generated probabilities effectively follow a uniform distribution, signifying the well-calibrated predictive models. 

\subsection{Daily cloud cover series at 3 p.m. }
This subsection briefly summarizes the results for the daily cloud cover series observed at 3 p.m., which are analogous, with the corresponding figures and tables provided in Sections S7 and S8 of the supplementary materials.

The largest value of the $\Lambda(\tau)$ statistic shown in Figure~S8, $\Lambda_{\text{max}} = 84.566$, occurs at $\hat{\tau} = 5109$ (December 30, 1978), exceeding the 95th empirical percentile (19.453) from Table~\ref{EmPercentile}, and indicating a similarly significant change in late 1978. The parameter estimates and their corresponding 95\% confidence intervals for the marginal mean models, with and without the estimated changepoint at $\hat{\tau} = 5109$, are summarized in Columns 4 and 5 of Table~S1 in the supplementary materials. The average of $|\hat{\Delta}_{k}|$, representing the absolute change in the logits of cumulative probabilities, is 0.644, which is slightly larger than the corresponding value for the daily cloud cover series observed at 9 a.m.

Additionally, the fitted marginal mean values during 1978 and 1979 are plotted in Figure~S7, along with the observed 3 p.m. daily cloud cover series.  The estimated categorical probabilities, with and without the estimated changepoint, are shown in Figure~S9.  Notably, Category 1 exhibited the largest change in estimated categorical probabilities; Categories 2 and 4 were also substantially affected, while Categories 3 and 5 were less affected. Finally, the PIT histogram in Figure~S10 suggests that the model provides an adequate fit to the 3 p.m. daily series.

Overall, for both daily cloud cover series observed at 9 a.m. and 3 p.m., the detected changepoints occur approximately one year prior to the documented changepoint, consistent with the findings of \citet{lu2012extended}, but identified with greater precision due to the daily data resolution.

\section{Discussion}\label{conclusion}
 
Motivated by the goal of accurately identifying structural changes in daily total cloud cover, this paper develops a changepoint test based on a maximally selected likelihood ratio statistic for ordinal categorical time series that exhibit seasonality and autocorrelation, modeled using a marginalized transition framework. Applying the proposed method to daily sky-cloudiness conditions at Fort St. John Airport, we successfully detect a changepoint that aligns with the findings of \citet{lu2012extended}, but with improved temporal precision at the daily level. In this study, the dependence structure of the series is modeled using a first-order Markov chain. The framework can be further extended by incorporating higher-order Markov chains or adopting seasonal Markov chain models to capture more complex temporal dependencies in real-world applications.

While the assumption of at most one changepoint (AMOC) simplifies the statistical framework, it may be unrealistic for longer time series, especially in high-frequency data such as daily or hourly observations. A natural extension of the current framework is to allow for the detection of multiple changepoints. Given that our marginalized transition model (MTM) supports likelihood-based inference, penalized likelihood approaches—such as the minimum description length (MDL) criterion or a modified Bayesian information criterion (BIC)—offer promising strategies for estimating multiple changepoints. This extension warrants further investigation in future research.

The primary limitation of our approach is computation time, which becomes more pronounced with larger datasets. For example,  applying the proposed $\Lambda_{\text{max}}$ test to 30 years of daily cloud cover data from Fort St. John Airport is computationally intensive. To efficiently search for the changepoint within the admissible set $\ell \leq \tau/n \leq h$, we employed a genetic algorithm (GA) \citep{davis2006structural, lu2010mdl, hewaarachchi2017homogenization}, specifically tailored to the $\Lambda_{\text{max}}$ test.  This approach substantially reduces the computational burden typically associated with exhaustive searches, making the procedure feasible even for large datasets. Details of the algorithm are provided in Section~S6 of the supplementary materials.

% By implementing the GA on a Mac Studio equipped with an M2 Max processor and 64 GB of RAM,  we reduced computation time from 13.4 hours to approximately 14 minutes. For the daily cloud cover series, the test statistic $\Lambda_{\text{max}} = 78.895$ occurs at $\hat{\tau} = 5024$ (October 6, 1978) for the 9 a.m. series, and $\Lambda_{\text{max}} = 77.353$ occurs at $\hat{\tau} = 5042$ (October 24, 1978) for the 3 p.m. series. These results closely align with those obtained through exhaustive search.   
% Nevertheless, further improvements in computational efficiency will be essential to ensure the method remains practical for large-scale or high-frequency applications.

 By implementing the GA on a Mac Studio equipped with an M2 Max processor and 64 GB of RAM,  we reduced computation time from 13.4 hours to approximately 11.0 minutes. For the daily cloud cover series, the test statistic $\Lambda_{\text{max}} = 79.961$ occurs at $\hat{\tau} = 5028$ (October 10, 1978) for the 9 a.m. series, and $\Lambda_{\text{max}} = 77.353$ occurs at $\hat{\tau} = 5042$ (October 24, 1978) for the 3 p.m. series. These results closely align with those obtained through exhaustive search.   
Nevertheless, further improvements in computational efficiency will be essential to ensure the method remains practical for large-scale or high-frequency applications.

\begin{acks}[Acknowledgments]
We would like to thank the anonymous reviewers for their constructive comments,
which significantly contributed to presenting the research in a better way.

High-Performance Computing resources provided by the High-Performance Research Computing (HPRC) core facility at Virginia Commonwealth University (\url{https://hprc.vcu.edu}) were used for conducting the research reported in this work.

Portions of this research were conducted with high performance computational resources provided by the Louisiana Optical Network Infrastructure (\url{http://www.loni.org}).

\end{acks}

%%%%%%%%%%%%%%%%%%%%%%%%%%%%%%%%%%%%%%%%%%%%%%
%% Funding information, if any,             %%
%% should be provided in the                %%
%% funding section.                         %%
%%%%%%%%%%%%%%%%%%%%%%%%%%%%%%%%%%%%%%%%%%%%%%
\begin{funding}
The Climate Research Division of the Atmospheric Science and Technology Directorate of Environment Canada is acknowledged for supporting QiQi Lu through the research contracts KM040-13-1319.

\end{funding}

\begin{supplement}
\stitle{Data and R code}
\sdescription{The data used in this study are available upon reasonable request from the authors. R and Rcpp code for analysis of case studies can be retrieved from \url{https://github.com/mli171/MTMCP}.}

%%%%%%%%%%%%%%%%%%%%%%%%%%%%%%%%%%%%%%%%%%%%%%
%% Supplementary Material, including data   %%
%% sets and code, should be provided in     %%
%% {supplement} environment with title      %%
%% and short description. It cannot be      %%
%% available exclusively as external link.  %%
%% All Supplementary Material must be       %%
%% available to the reader on Project       %%
%% Euclid with the published article.       %%
%%%%%%%%%%%%%%%%%%%%%%%%%%%%%%%%%%%%%%%%%%%%%%

Supplementary materials include the following sections: Section 1: Detailed derivation of Quasi-Newton's Method for MTM, Section 2: Detailed Derivation of Quasi-Newton’s Method for MTM, Section 3: Detailed Derivation of $\delta_{tk}$ for MTM, Section 4: Individual model coefficients inference, Section 5: Model goodness of fit, Section 6: Island model genetic algorithm for single changepoint detection, Section 7: Supplementary Figures, Section 8: Supplementary Tables.

\setcounter{section}{0}
\setcounter{table}{0}
\setcounter{figure}{0}

\renewcommand{\thesection}{S\arabic{section}}
\renewcommand{\thetable}{S\arabic{table}}
\renewcommand{\thefigure}{S\arabic{figure}}

\section{Detailed Parameter Estimation Procedure via the Quasi-Newton algorithm}
We update $\bm{\theta}$ and $\bm{\xi}$ with the initial values $\bm{\theta}^{(0)}$ and $\bm{\xi}^{^{(0)}}$ iteratively.  At iteration $i$ with $\bm{\theta}^{(i)}$ and $\bm{\xi}^{^{(i)}}$,\\
\underline{Step 1}: Given $\bm{\xi}^{^{(i)}}$, let $\tilde{\bm{\theta}} = \bm{\theta}$ and find the MLE estimate of $\tilde{\bm{\theta}}$, denoted as $\tilde{\bm{\theta}}^{*}$ The log-likelihood function, $\log{L(\tilde{\bm{\theta}}, \bm{\xi}^{(i)})}$, is maximized using a new Quasi-Newton algorithm with the initial value $\tilde{\bm{\theta}}^{(0)}=\bm{\theta}^{(i)}$. The updating equation at the $j$th iteration is given by
\begin{equation*} \label{eqthetaupdate}
    \tilde{\bm{\theta}}^{(j+1)} = \tilde{\bm{\theta}}^{(j)} - \bigg(I_{\tilde{\bm{\theta}}}(\tilde{\bm{\theta}}^{(j)},\bm{\xi}^{^{(i)}};\bm{Y})\bigg)^{-1}\frac{\partial\text{log}L(\tilde{\bm{\theta}}^{(j)}, \bm{\xi}^{^{(i)}};\bm{Y})}{\partial\tilde{\bm{\theta}}^{(j)}},
\end{equation*}
where
\[
    I_{\tilde{\bm{\theta}}}(\tilde{\bm{\theta}}^{(j)},\bm{\xi}^{(i)};\bm{Y}) = \sum_{t=1}^{n}\bigg(\frac{\partial\text{log}L(\tilde{\bm{\theta}}^{(j)},\bm{\xi}^{(i)};y_{t})}{\partial\tilde{\bm{\theta}}^{(j)}}\bigg)\bigg(\frac{\partial\text{log}L(\tilde{\bm{\theta}}^{(j)},\bm{\xi}^{(i)};y_{t})}{\partial\tilde{\bm{\theta}}^{(j)}}\bigg)^{'}
\]
is the consistent estimator of the information matrix for $\tilde{\bm{\theta}}$.
To obtain $\tilde{\bm{\theta}}^{*}$, the update of $\tilde{\bm{\theta}}$ is repeated until the convergence criterion 
\[
\sqrt{(\tilde{\bm{\theta}}^{(j+1)}-\tilde{\bm{\theta}}^{(j)})^{'}(\tilde{\bm{\theta}}^{(j+1)}-\tilde{\bm{\theta}}^{(j)})} \leq \epsilon
\]
is satisfied. 
Then we set $\bm{\theta}^{(i+1)}=\tilde{\bm{\theta}}^{*}$ and most importantly, we have to check the convergence for both $\bm{\theta}^{(i+1)}$ and $\bm{\xi}^{(i)}$ jointly using (9) with $\hat{\bm{\phi}}^{new}=(\bm{\theta}^{(i+1)}, \bm{\xi}^{(i)})$ and $\hat{\bm{\phi}}^{old}=(\bm{\theta}^{(i)}, \bm{\xi}^{(i)})$. If the convergence is reached, then this $\hat{\bm{\phi}}^{new}$ is our final estimate of $\bm{\phi}$. Otherwise, we need to update $\bm{\xi}$ in Step 2.\\
\underline{Step 2}: Given $\bm{\theta}^{^{(i+1)}}$, let $\tilde{\bm{\xi}}= \bm{\xi}$ and find the MLE estimate of $\tilde{\bm{\xi}}$, denoted as $\tilde{\bm{\xi}}^{*}$. Here, $\log{L(\bm{\theta}^{(i+1)}, \tilde{\bm{\xi}})}$ is maximized using another new Quasi-Newton algorithm with the initial value $\tilde{\bm{\xi}}^{(0)}=\bm{\xi}^{(i)}$. The updating equation at the $j$th iteration is 
\begin{equation*} \label{eqxiupdate}
	\tilde{\bm{\xi}}^{(j+1)} = \tilde{\bm{\xi}}^{(j)} - \bigg(I_{\tilde{\bm{\xi}}}(\bm{\theta}^{(i+1)},\tilde{\bm{\xi}}^{(j)};\bm{Y})\bigg)^{-1}\frac{\partial\text{log}L(\bm{\theta}^{(i+1)}, \tilde{\bm{\xi}}^{(j)};\bm{Y})}{\partial\tilde{\bm{\xi}}^{(j)}},
\end{equation*}
where 
\begin{align*}
    I_{\tilde{\bm{\xi}}}(\bm{\theta}^{(i+1)},\tilde{\bm{\xi}}^{(j)};\bm{Y}) & = \sum_{t=1}^{n}\bigg(\frac{\partial\text{log}L(\bm{\theta}^{(i+1)},\tilde{\bm{\xi}}^{(j)};y_{t})}{\partial\tilde{\bm{\xi}}^{(j)}}\bigg)\bigg(\frac{\partial\text{log}L(\bm{\theta}^{(i+1)},\tilde{\bm{\xi}}^{(j)};y_{t})}{\partial\tilde{\bm{\xi}}^{(j)}}\bigg)^{'}
\end{align*}
is the consistent estimator of the information matrix for $\tilde{\bm{\xi}}$. 
Similarly, the update of $\bm{\xi}$ is repeated until convergence of
\[
\sqrt{(\tilde{\bm{\xi}}^{(j+1)}-\tilde{\bm{\xi}}^{(j)})^{'}(\tilde{\bm{\xi}}^{(j+1)}-\tilde{\bm{\xi}}^{(j)})}\leq \epsilon
\]
is met, and then $\tilde{\bm{\xi}}^{*}$ is obtained.
Now we set $\bm{\xi}^{(i+1)}=\tilde{\bm{\xi}}^{*}$ and check the joint convergence in (9) for $\bm{\theta}^{(i+1)}$ and $\bm{\xi}^{(i+1)}$.  This time,  $\hat{\bm{\phi}}^{new}=(\bm{\theta}^{(i+1)}, \bm{\xi}^{(i+1)})$ and $\hat{\bm{\phi}}^{old}=(\bm{\theta}^{(i+1)}, \bm{\xi}^{(i)})$. Steps 1 and 2 are repeated until the joint convergence of $\bm{\theta}^{(i+1)}$ and $\bm{\xi}^{(i+1)}$  in (9) is met. Typically, joint convergence is reached after only one or two iterations.

%\newpage

\section{Detailed Derivation of Quasi-Newton's Method for MTM\label{appendix:MTMNewton}}
Provided the log-likelihood function in (8), the gradient vector $\frac{\partial\text{log}L(\bm{\phi},\bm{Y})}{\partial\bm{\phi}}$ are given in this section. Recall the parameter vector $\bm{\phi}=(\bm{\theta}, \bm{\xi})$, where $\bm{\theta}=(\alpha_{1}, \ldots, \alpha_{K-1}, B_{1}, \ldots, B_{K-1}, \\ D_{1}, \ldots, D_{K-1}, \beta_{1}, \ldots, \beta_{K-1}, \Delta_{1}, \ldots, \Delta_{K-1})'$ and $\bm{\xi}=(\xi_{ab})_{a,b=1}^{K-1}$.

Let $\theta_{p}$ represent the category-specific parameter contained in $\bm{\theta}$ for category $p$, where $p=1,\ldots,K-1$. The forms of derivatives for gradient follow
\begin{align*}
	\frac{\partial\text{log} L_{1}(\bm{\theta})}{\partial\theta_{p}} =& \frac{y_{1,1}}{\gamma_{1,1}}\frac{\partial\gamma_{1,1}}{\partial\theta_{p}}+\sum_{k=2}^{K-1}\frac{y_{1,k}}{\gamma_{1,k}-\gamma_{1,k-1}}\bigg(\frac{\partial\gamma_{1,k}}{\partial\theta_{p}} - \frac{\partial\gamma_{1,k-1}}{\partial\theta_{p}}\bigg) +\frac{y_{1,K}}{1-\gamma_{1,K-1}}\bigg(-\frac{\partial\gamma_{1,K-1}}{\partial\theta_{p}}\bigg),\\
	\frac{\partial\text{log} L_{2}(\bm{\theta}, \bm{\xi})}{\partial\theta_{p}} =& \sum_{t=2}^{n}\sum_{k=1}^{K-1}\Big((y_{tk}-p_{t,jk})\frac{\partial\delta_{t,k}}{\partial\theta_{p}}\Big),\\
    \frac{\partial\text{log} L_{2}(\bm{\theta}, \bm{\xi})}{\partial\xi_{ab}} =& \sum_{t=2}^{n}\bigg\{\sum_{k=1}^{K-1}(y_{tk}-p_{t,jk})\frac{\partial\delta_{t,k}}{\partial\xi_{ab}}+(y_{ta}-p_{t,ja})y_{t-1,b}\bigg\},
\end{align*}
where $\frac{\partial\gamma_{t,k}}{\partial\theta_{p}}=\gamma_{t,k}(1-\gamma_{t,k})\frac{\partial\eta_{t,k}}{\partial\theta_{p}}$ and $\frac{\partial\eta_{t,k}}{\partial\theta_{p}} = x_{t}$ if $k=p$, $\frac{\partial\eta_{t,k}}{\partial\theta_{p}} = 0$ otherwise. $x_{t}$ denotes the covariate associated with $\theta$ at time $t$.

To compute the derivatives for the gradient and the information matrix, $I_{e}(\bm{\theta},\bm{\xi};\bm{Y})$, we also need derivatives of $\delta_{t,k}$ with respect to $\bm{\theta}$, $\frac{\partial\delta_{t,k}}{\partial\theta_{p}}$, and with respect to $\bm{\xi}$, $\frac{\partial\delta_{t,k}}{\partial\xi_{ab}}$. These values can be obtained as the solutions to the following system of linear equations,
\begin{align*}
	\frac{\partial\delta_{t,1}}{\partial\theta_{p}}\sum_{j=1}^{K}\frac{\partial p_{t,jk}}{\partial\delta_{t,1}}\pi_{t-1,j}+\ldots+\frac{\partial\delta_{t,K-1}}{\partial\theta_{p}}\sum_{j=1}^{K}\frac{\partial p_{t,jk}}{\partial\delta_{t,K-1}}\pi_{t-1,j}=&\frac{\partial\pi_{t,k}}{\partial\theta_{p}}-\sum_{j=1}^{K}p_{t,jk}\frac{\partial\pi_{t-1,j}}{\partial\theta_{p}},\\
	\frac{\partial\delta_{t,1}}{\partial\xi_{ab}}\sum_{j=1}^{K}\frac{\partial p_{t,jk}}{\partial\delta_{t,1}}\pi_{t-1,j}+\ldots+\frac{\partial\delta_{t,K-1}}{\partial\xi_{ab}}\sum_{j=1}^{K}\frac{\partial p_{t,jk}}{\partial\delta_{t,K-1}}\pi_{t-1,j}
	=&-\sum_{j=1}^{K}\frac{\partial p_{t,jk}}{\partial\xi_{ab}}\pi_{t-1,j},
\end{align*}
where 
\begin{align*}
	\frac{\partial p_{t,jk}}{\partial\delta_{tk'}} =&
			\begin{cases}
            p_{t,jk} - p_{t,jk}p_{t,jk'},  & \text{if $k=k', k=1,\ldots,K-1,k'=1,\ldots,K-1$;} \\
            - p_{t,jk}p_{t,jk'},  & \text{if $k\neq k', k=1,\ldots,K,k'=1,\ldots,K-1$;}
        \end{cases} \\
    \frac{\partial p_{t,jk}}{\partial\xi_{ab}}=&
    		\begin{cases}
            	p_{t,jk} - p_{t,jk}p_{t,ja},  & \text{if $b=j$, $k=a$, $k=1,\ldots,K-1$, $j=1,\ldots,K-1$;} \\
            	- p_{t,jk}p_{t,ja},  & \text{if $b=j$, $k\neq a$, $k=1,\ldots,K$, $j=1,\ldots,K-1$;} \\
	            0,  & \text{if $b\neq j$, $k=1,\ldots,K-1$, $j=1,\ldots,K-1$.}
            \end{cases}
\end{align*}
We could solve for $\frac{\partial\delta_{t,k}}{\partial\theta_{p}}$ and $\frac{\partial\delta_{t,k}}{\partial\xi_{ab}}$ using these $K-1$ equations.

%\newpage
\section{Detailed Derivation of $\delta_{tk}$ for MTM \label{appendix:MTMdelta}}
\noindent Let $\vec{f}(\vec{\delta}_{t})=\{f_{1}(\vec{\delta}_{t}), \ldots, f_{K-1}(\vec{\delta}_{t})\}$, where $\vec{\delta}_{t} = \{\delta_{t,1}, \ldots, \delta_{t,K-1}\}$ and $f_{k}(\vec{\delta}_{t})=\sum_{j=1}^{K}p_{t,jk}\pi_{t-1,j}-\pi_{t,k}$. The conditional probabilities $\text{P}(Y_{t}=k|Y_{t-1}=j)$ are denoted as $p_{t,jk}$. Then, given the values of model parameters, $\bm{\theta}$ and $\bm{\xi}$, we could use the Newton-Raphson algorithm to numerically estimate $\delta_{t,k}$.
\begin{equation*}
	\vec{\delta}_{t}^{(i+1)}=\vec{\delta}_{t}^{(i)}-\bigg(\frac{\partial\vec{f}(\vec{\delta}_{t})}{\partial\vec{\delta}_{t}} \bigg)^{-1} \vec{f}(\vec{\delta}_{t}),
\end{equation*}
where the $\frac{\partial\vec{f}(\vec{\delta}_{t})}{\partial\vec{\delta}_{t}}$ is a $(K-1)$ by $(K-1)$ matrix and each element in this matrix are given as follows,
\begin{equation*}
    \frac{\partial f_{k}(\vec{\delta}_{t})}{\partial\delta_{t,k'}}=
    \begin{cases}
        \sum_{j=1}^{K}p_{t,jk}(1-p_{t,jk})\pi_{t-1,j}, & \text{if}\quad k=k';\\
        -\sum_{j=1}^{K}p_{t,jk}p_{t,jk'}\pi_{t-1,j}, & \text{if}\quad k\neq k'.\\
    \end{cases}
\end{equation*}

%\newpage 
\section{Individual model coefficients inference}

For large samples, the distribution of the maximum-likelihood estimator is approximately normal with little bias, as shown in Figures~S2 and S3 for marginal mean regression parameters ($\bm{\theta}$), and in Figures~S4 and S5 for dependence model parameters ($\bm{\xi}$). We can find the variances and covariances matrix for $\bm{\theta}$ and $\bm{\xi}$ from the estimated second partial derivatives of Hessian matrix $I_{\phi}(\bm{\hat{\phi}},\bm{Y})$ in (10) evaluated at the maximum-likelihood estimates,
\begin{equation*}
    \text{var}(\bm{\hat{\beta}}) = - I_{\phi}(\bm{\hat{\phi}},\bm{Y})^{-1}.
\end{equation*}
Then the square roots of the diagonal elements of matrix $\text{var}(\bm{\hat{\beta}})$ are the large sample standard errors of the model coefficients. The Wald inference could be conducted for each parameter estimates $\hat{\phi}$ \citep{montgomery2021introduction} with the test statistic
\begin{equation*}\label{waldtest}
    Z_{0} = \frac{\hat{\phi}}{\text{se}(\hat{\phi})},
\end{equation*}
with the reference standard normal distribution. Related, the $100(1-\alpha)$ percent confidence interval on $\phi$ can be constructed as
\begin{equation*}\label{waldCI}
    \hat{\phi} - Z_{\alpha/2}\text{se}(\hat{\phi}) \leq \phi \leq \hat{\phi} + Z_{1-\alpha/2}\text{se}(\hat{\phi}),
\end{equation*}
where $Z_{\alpha/2}$ and $Z_{1-\alpha/2}$ are related reference distribution critical values.

%\newpage
\section{Model goodness of fit}
The probability integral transform (PIT) histogram was introduced by \cite{czado2009predictive} as a graphical tool to assess the goodness of fit when modeling categorical or count data with discrete predictive distributions \citep{czado2009predictive, kolassa2016evaluating, baran2021machine, jia2023latent}. For a well-fitted model, one anticipates statistical consistency between the predictive distribution and the observed ordinal time series, reflected by the PIT histogram exhibiting a pattern akin to the standard uniform distribution. The observed deviations from uniformity indicate potential model deficiencies, such as issues related to parameter estimation or dispersion \citep{czado2009predictive}. 

To construct the PIT histogram, we define the conditional cumulative probability of $Y_{t}$ as
\begin{align*}
    \mathbb{P}_{t}(y) &= P(Y_{t}\leq y| Y_{t-1}=y_{t-1}, \ldots, Y_{1}=y_{1})\\
    &= P(Y_{t}\leq y| Y_{\{t-1\}}), \quad y\in\{1,\ldots\,K\},
\end{align*}
where $Y_{\{t-1\}}=\{Y_{t-1}, \ldots, Y_{1}\}$ represents all the past values.  
Then, the nonrandomized mean PIT values can be calculated as
\begin{equation*}
    \bar{F}(u) = \frac{1}{n} \sum_{t=1}^{n}F_{t}(u|y_{t}), \quad u\in [0,1],
\end{equation*}
where
\begin{equation*}
F_{t}(u|y) = \begin{cases}
0, & \text{if} \ u\leq \mathbb{P}_{t}(y-1), \\
\frac{u-\mathbb{P}_{t}(y-1)}{\mathbb{P}_{t}(y)-\mathbb{P}_{t}(y-1)}, & \text{if} \ \mathbb{P}_{t}(y-1) < u < \mathbb{P}_{t}(y), \\
1, & \text{if} \ u \geq \mathbb{P}_{t}(y),
\end{cases}
\end{equation*}
is based on $\mathbb{P}_{t}(y)$ and the observed data. In practice, one can substitute $\mathbb{P}_{t}(y)$ with any estimator $\hat{\mathbb{P}}_{t}(y)$ according to the fitted models. For our MTM model, $\hat{\mathbb{P}}_{t}(y)$ can be obtained by aggregating the transition probabilities,
\begin{equation*}
    \hat{\mathbb{P}}_{t}(y) = \sum_{k=1}^{y}\hat{E}[I_{\{Y_{t}=k\}}|Y_{\{t-1\}}]
    =\sum_{k=1}^{y}\hat{P}(Y_{t}=k|Y_{\{t-1\}})
    =\sum_{k=1}^{y}\hat{P}(Y_{t}=k|Y_{t-1}=y_{t-1}),
\end{equation*}
where $\hat{P}(Y_{t}=k|Y_{t-1}=y_{t-1})$ is the predicted one-step transition probability and can be obtained from the dependence model in (6).

The PIT histogram with $H$ bins is defined as a histogram with the heights $\bar{F}(h/H) - \bar{F}((h-1)/H)$ for equally spaced bins $h = 1,\ldots,H$. Note that the selection of $H$ is arbitrary and should be based on the time series sample size $n$.

%\newpage
\section{Island model genetic algorithm for single changepoint detection}

We now propose using the island model genetic algorithm (GA) as an efficient alternative to the computationally demanding $\Lambda_{\text{max}}$ test for changepoint detection. Initially introduced by \cite{jh1975adaptation} and inspired by principles of natural selection, evolution, gene mutation, and population genetics, GA intelligently searches for the optimal solution without requiring exhaustive model evaluations, making it well-suited for time series with large sample sizes.

The algorithm begins with an initial population of potential changepoint locations represented as individuals. In each generation, individuals with higher log-likelihood function values have a greater chance of being selected as parents. Through a ``crossover'' procedure, child chromosomes inherit information from their parents for the next generation. Additionally, random changes in the child's chromosome, mimicking gene mutation, occur with a low probability to avoid getting stuck in local optimal solutions.
Using the steady-state approach, less fit individuals are replaced by newly produced offspring in constructing the new generation. As the population evolves, newer generations tend to include individuals with better fitness. The algorithm iterates until it converges on an individual that maximizes the likelihood function, signifying an approximate optimal solution.

The island model GA implementation for the at most one changepoint (AMOC) problem represents each individual as a changepoint location $\tau$ selected from the truncated admissible changepoint set, $\ell \leq \tau/n \leq h$.
We start with an initial population of $n_{p}=100$ randomly generated individuals. Fitness is evaluated using the log-likelihood functions defined in (6), with computations parallelized across 10 threads. In each generation, parent selection is performed using the linear ranking method, which is used to form chromosome pairs for crossover operator. The selection probability is determined by fitness rank, with the fittest individual having the highest rank ($n_{p}-1$) and the least fit individual having rank 0. It is worth noting that the father in each pair always has the better-fit chromosome. Then, the crossover operator with a probability of $p_{c}=0.9$ generates offspring for the next generation by averaging the parents' changepoint locations and rounding to the nearest integer or directly copying the father's chromosome if no child is produced.

Following crossover, the child chromosome undergoes a mutation with a low probability of $p_{m}=0.1$. If a mutation occurs, the changepoint location in the child chromosome can be randomly selected from the parameter space. During each population, we excluded the previously evaluated chromosomes of every individual to avoid redundant fitness function evaluations. This mutation operator ensures that no solution in the admissible changepoint set has a zero probability of being considered, thereby enhancing the algorithm's exploration of the solution space. 

The steady-state replacement method \citep{davis1991handbook} is used to form a new generation, where only one pair of parents' chromosomes is sampled to produce one child during the crossover. Duplicate children are checked and discarded to prevent repeated evaluations of the same log-likelihood function and avoid premature convergence to local optima. Consequently, only one individual in the current generation is replaced by a new child after crossover and mutation.

Migrations in the GA improve convergence speed and population diversity \citep{davis1991handbook}. The population is divided into $N_{I}=5$ islands, each with 20 individuals per generation. During migration, the least-fit individual from an island $j$ is replaced by the best-fit individual randomly selected from another island $j'$. This prevents premature convergence and promotes efficient exploration of the solution space. To reduce computational time, evolution on the five islands is performed in parallel.

Our island model GA uses a convergence criterion based on the approach described in \cite{davis2006structural}, where if the overall best-fit individual remains unchanged for $M_{c}=4$ consecutive migrations or the total migrations exceed the predetermined maximum number of migrations, $M_{max}=15$, the search process stops. The best-fit individual is considered the optimal solution.

%\newpage
\section{Supplementary Figures\label{figs}}
~

\begin{figure*}[h!]
\centering
\includegraphics[scale=0.33]{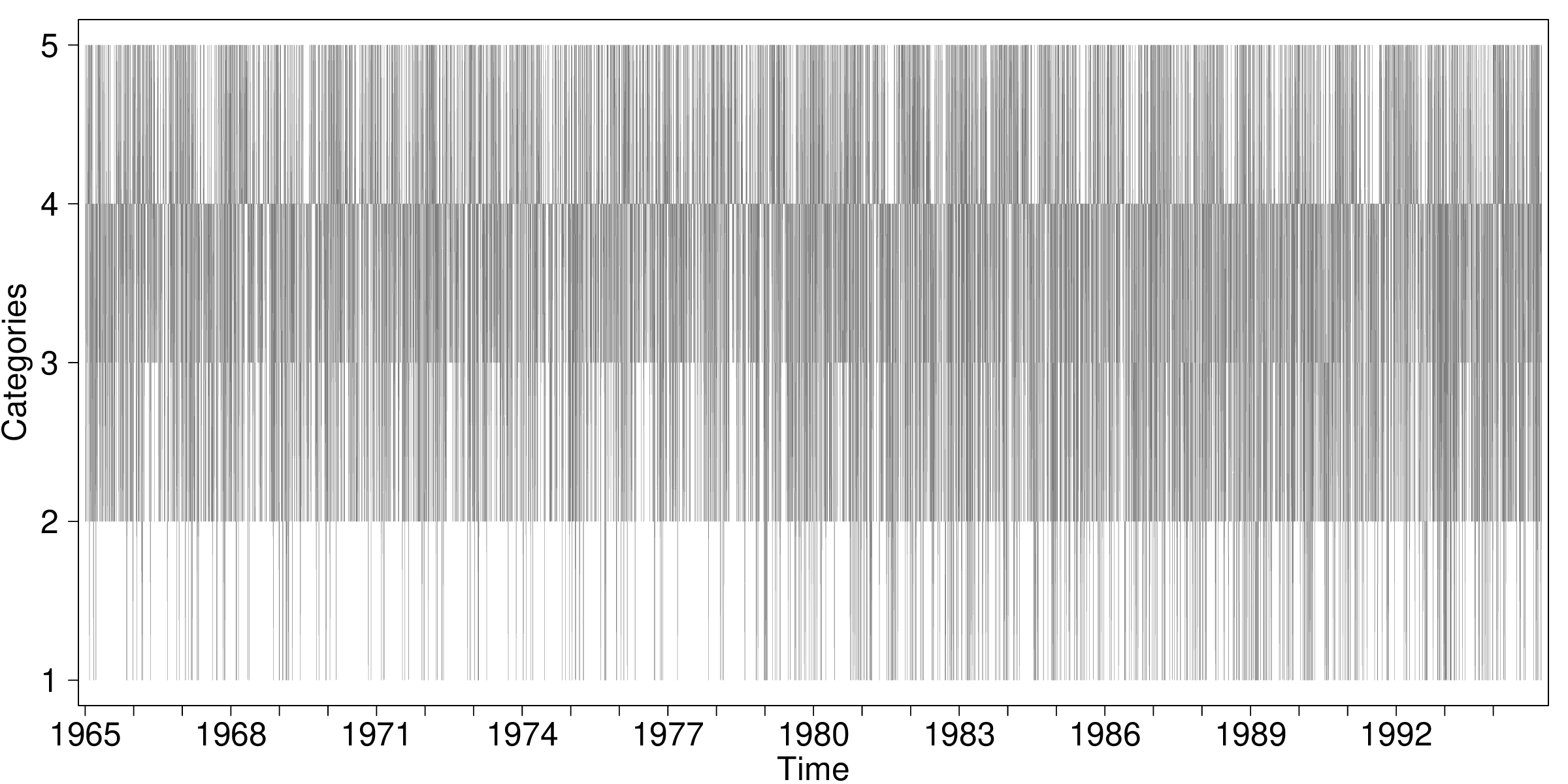}
\caption{Daily sky-cloudiness condition series observed at 3 p.m. with $K=5$ categories at Fort St. John Airport from 1965 to 1994.}
\end{figure*}    

\newpage

\begin{figure*}[h!]
\centerline{\includegraphics[scale=0.35]{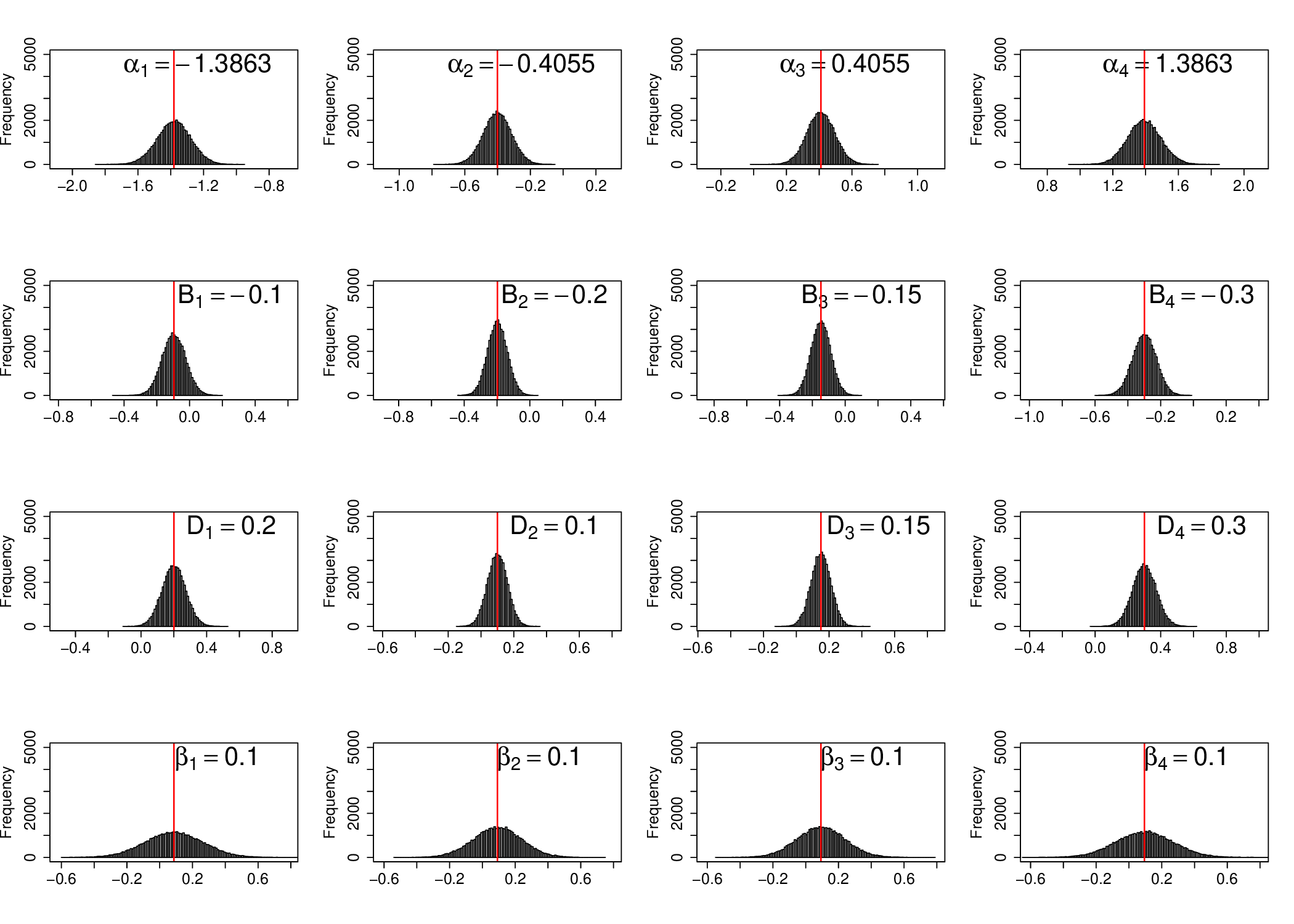}}
\caption{Histograms for marginal mean model parameter estimation with $K=5$, $n=3650$ and $T=365$.}
\end{figure*}

\begin{figure*}[h!]
\centering
\includegraphics[scale=0.35]{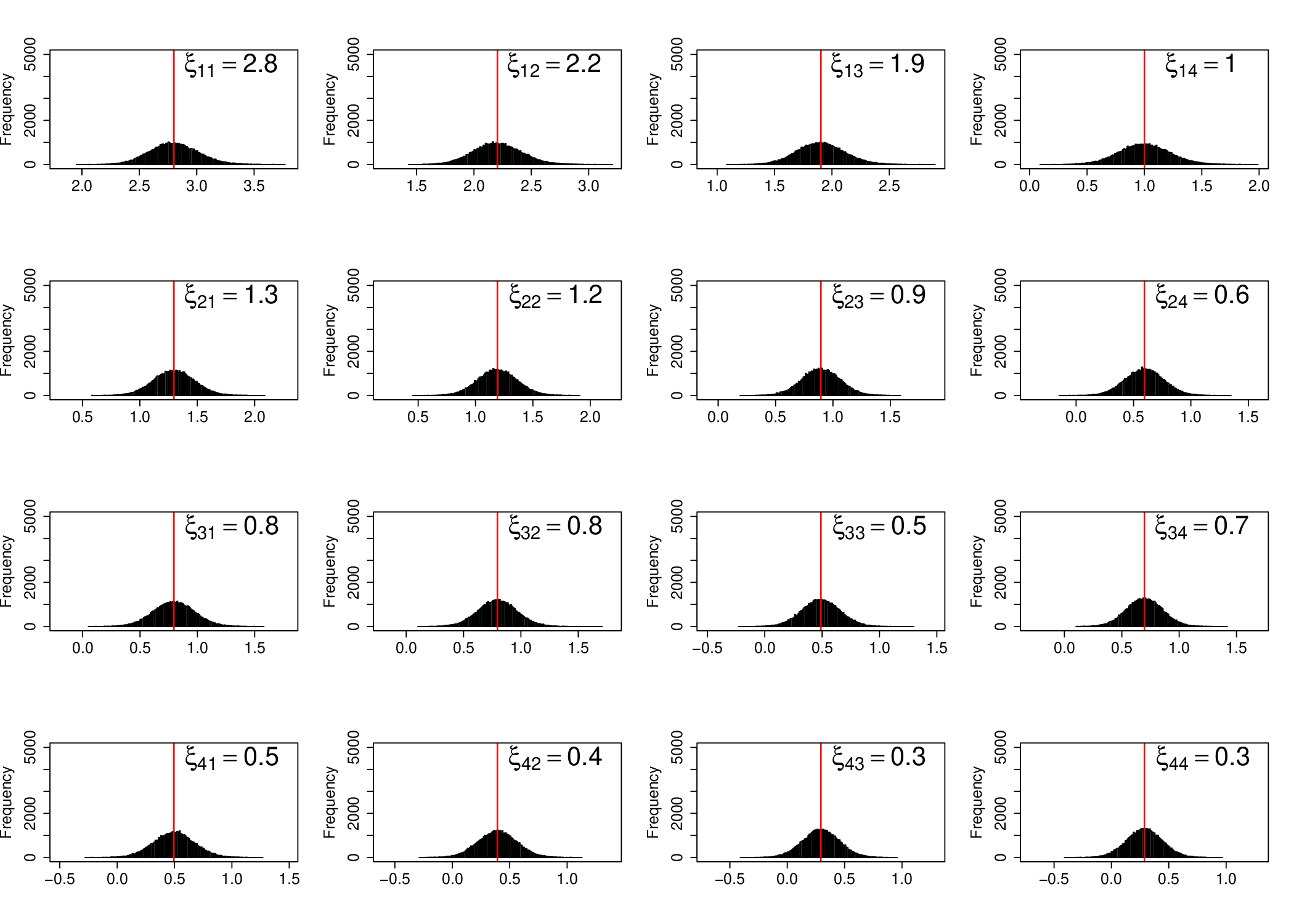}
\caption{Histograms for dependence model parameter estimation with $K=5$, $n=3650$ and $T=365$.}
\end{figure*}

\newpage

% Here, I only use .png file since it won't affect the compiling speed that much. I also upload the real .eps file to figs (large size files).
\begin{figure*}[h!]
\centering
\includegraphics[scale=0.075]{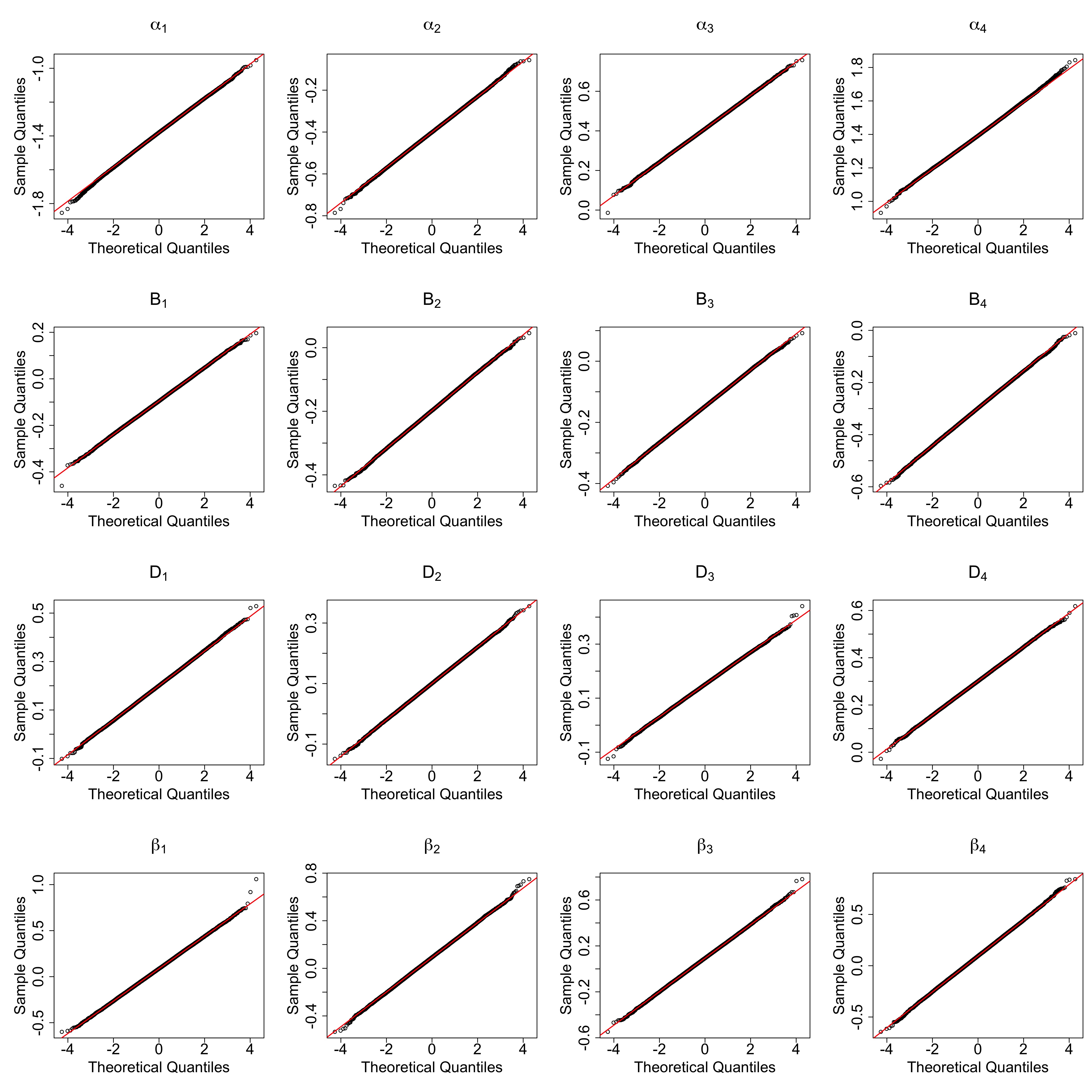}
\caption{QQ plots of marginal mean model parameter estimation with $K=5$, $n=3650$ and $T=365$ \label{fig3}}
\end{figure*}

\newpage

\begin{figure*}[h!]
\centering
\includegraphics[scale=0.065]{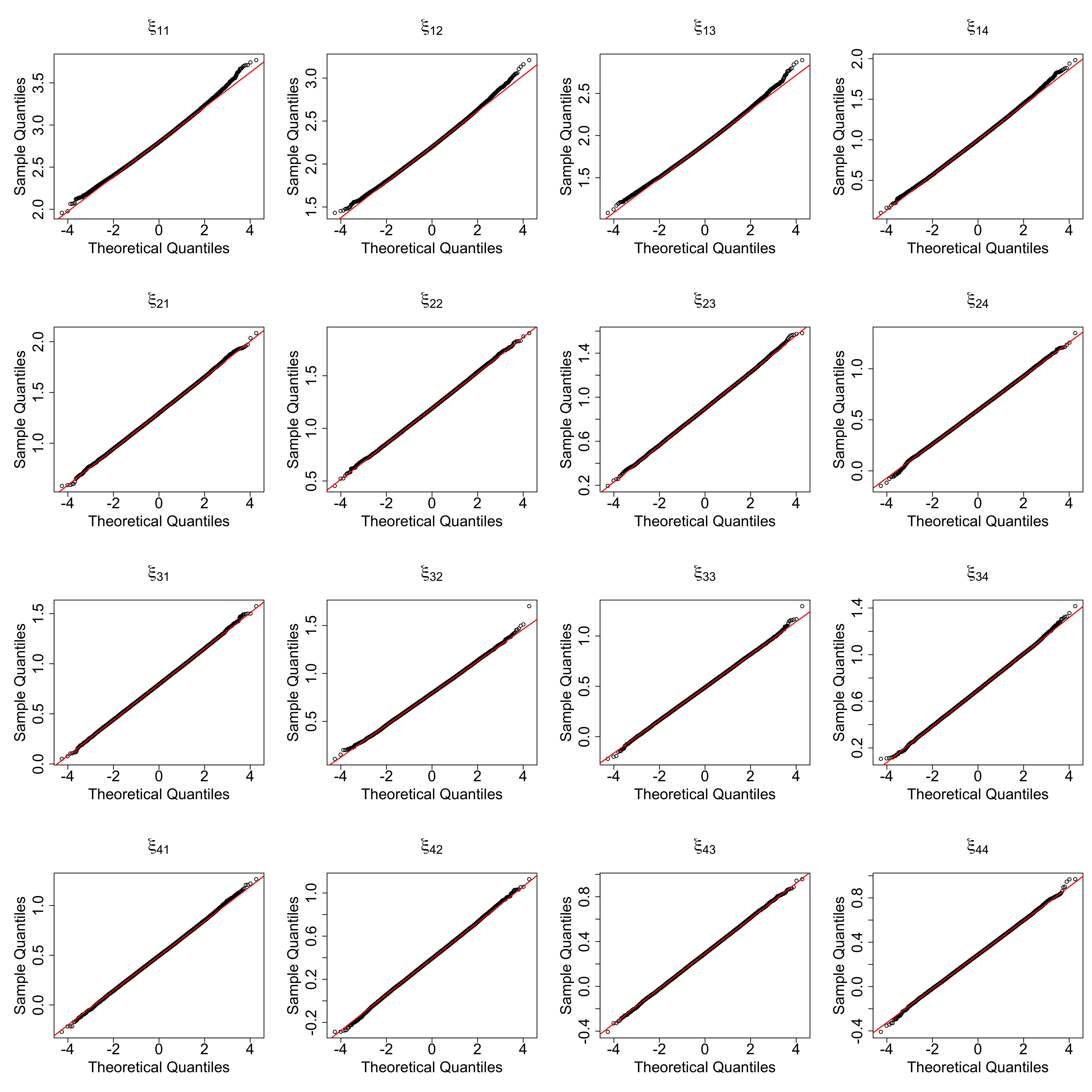}
\caption{QQ plots of dependence model parameter estimation with $K=5$, $n=3650$ and $T=365$}
\end{figure*}

%\newpage 
 
\begin{figure*}[h!]
\centerline{\includegraphics[scale=0.257]{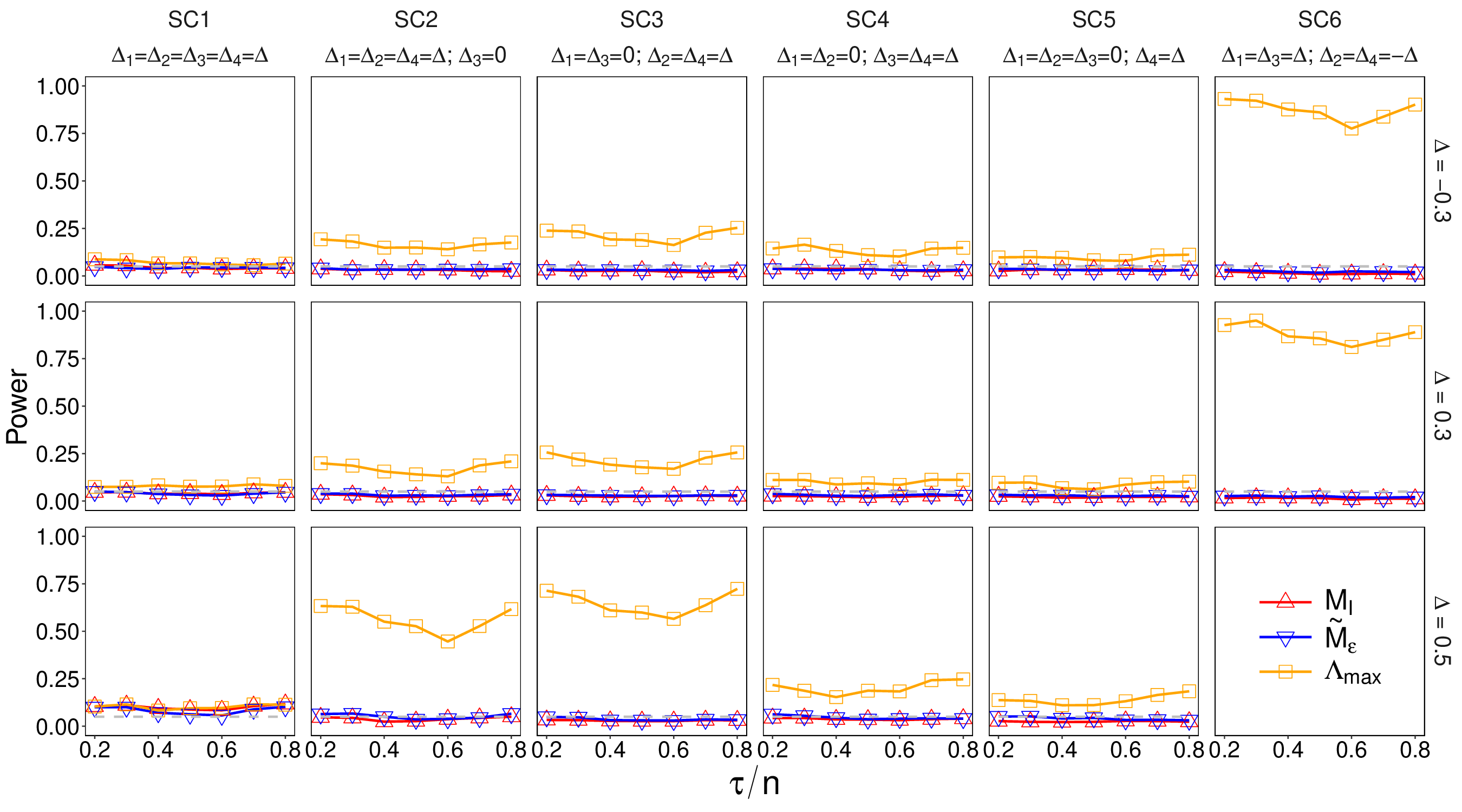}}
\caption{Estimated detection power of the $\Lambda_{\text{max}}$, $M_{I}$, and $\tilde{M}_{\epsilon}$ tests for the marginal model with trends under strong positive autocorrelation, plotted as a function of $\tau/n$ for three levels of $\Delta$.}
%\caption{Detection power comparison under the strong positive correlation structure with dependence parameter set $\bm{\xi}_{1}$ for Model 2.}
\end{figure*}

% \newpage
% \begin{figure*}[h!]
% \centerline{\includegraphics[scale=0.3]{figs/MedPosComparison_model1.eps}}
% \caption{Medium positive correlation power comparison for Model 1.}
% \end{figure*}

% \begin{figure*}[h!]
% \centerline{\includegraphics[scale=0.3]{figs/MedPosComparison_model2.eps}}
% \caption{Medium positive correlation power comparison for Model 2.}
% \end{figure*}

% \newpage
% \begin{figure*}[h!]
% \centerline{\includegraphics[scale=0.3]{figs/IndComparison_model1.eps}}
% \caption{Independent correlation power comparison for Model 1.}
% \end{figure*}

% \begin{figure*}[h!]
% \centerline{\includegraphics[scale=0.3]{figs/IndComparison_model2.eps}}
% \caption{Independent correlation power comparison for Model 2.}
% \end{figure*}

\vspace{0.4 in}

\begin{figure*}[h!]
\centerline{\includegraphics[scale=0.33]{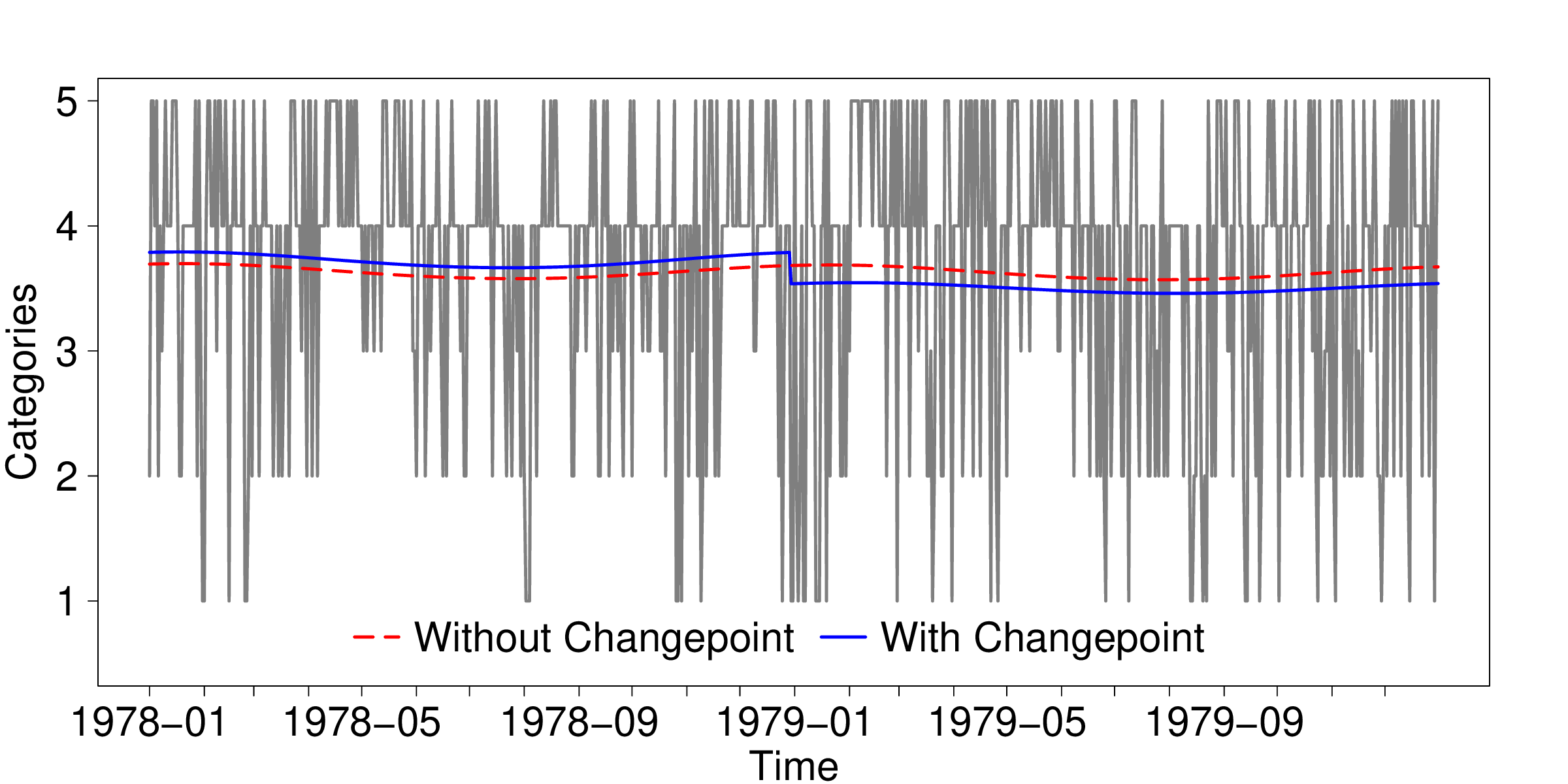}}
\caption{Daily sky-cloudiness conditions series observed at 3 p.m. at Fort St. John Airport in 1978 and 1979, with the estimated mean and the changepoint (December 30, 1978).}
\end{figure*}

%\newpage

\begin{figure*}[!h]
\centerline{\includegraphics[scale=0.31]{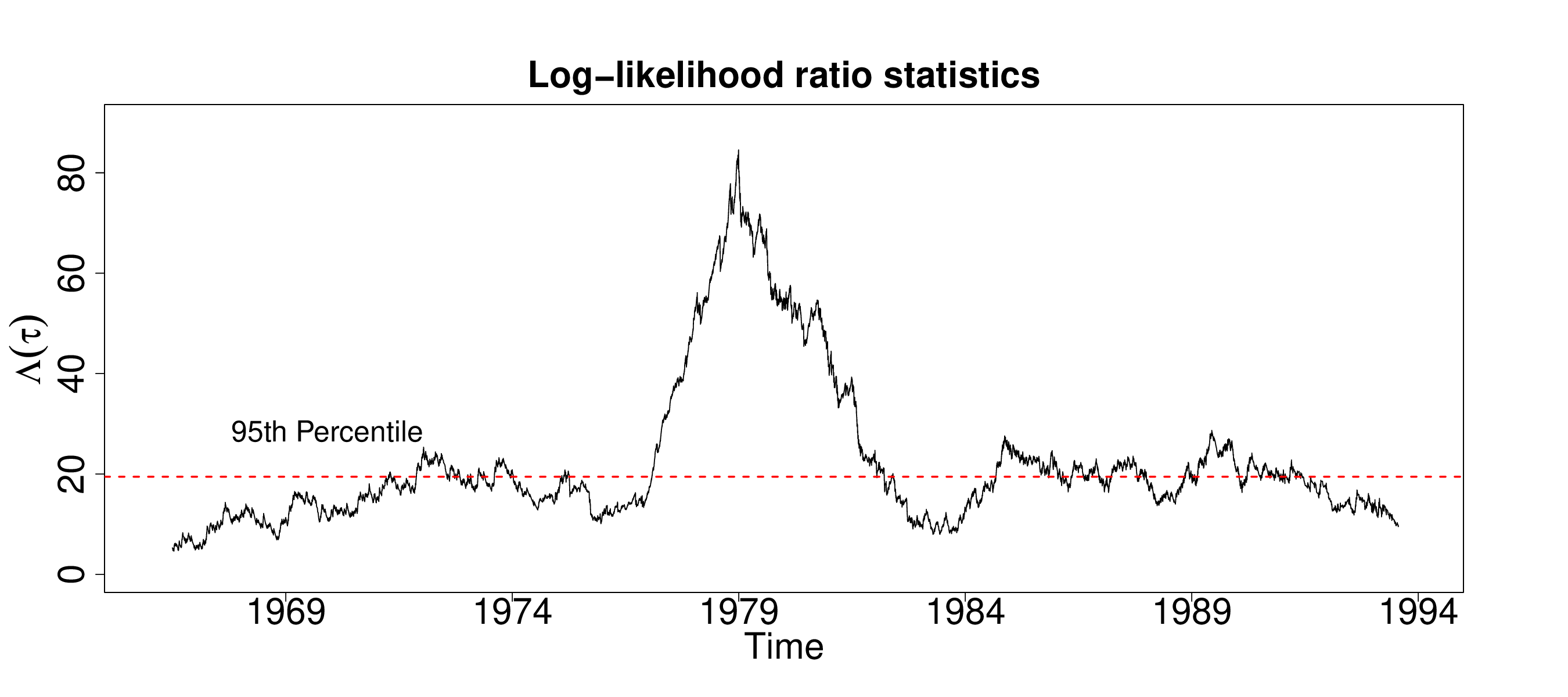}}
\caption{Log-likelihood ratio statistics $\Lambda(\tau)$ for daily Sky-Cloudiness Data observed at 3 p.m. at Fort St. John Airport.}
\end{figure*}

\vspace{1 in}

\begin{figure*}[!h]
\centerline{\includegraphics[scale=0.31]{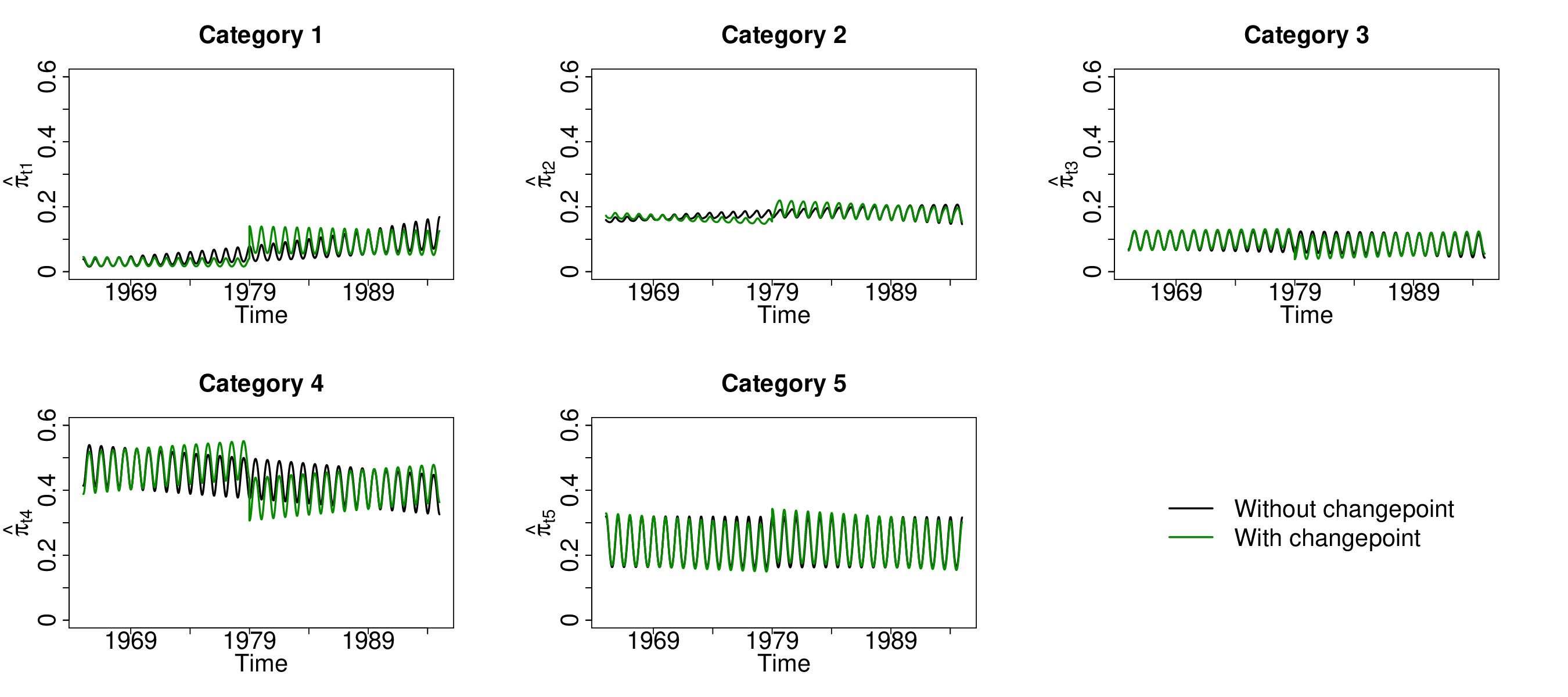}}
\caption{Estimated categorical probabilities $\hat{\pi}_{t,k}$ without a changepoint (black) and with a changepoint (green) across five categories for daily sky-cloudiness data observed at 3 p.m. at Fort St.~John Airport.}
\end{figure*}

\newpage

\begin{figure*}[!ht]
\centerline{\includegraphics[scale=0.32]{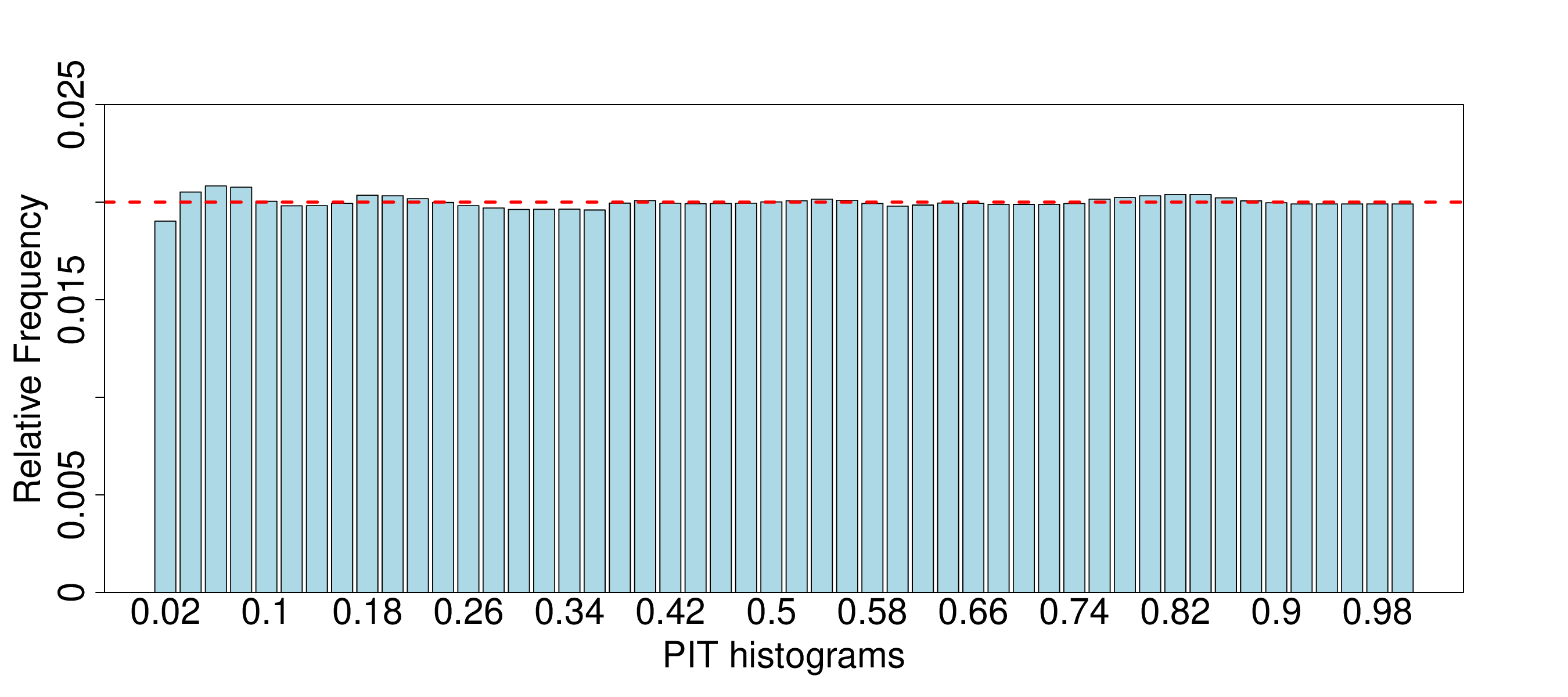}}
\caption{PIT histogram for the marginal mean model with the estimated changepoint for daily sky-cloudiness conditions observed at 3 p.m. at Fort St. John Airport.}
\end{figure*}

% \newpage

% \begin{figure*}[!h]
% \centerline{\includegraphics[scale=0.31]{figs/ResidualDiagnostic_Hour15.eps}}
% \caption{Sample autocorrelations of the estimated one-step ahead prediction residuals $e_{tk}$ for Category $k = 1, \ldots, 4$ with 95\% pointwise confidence bands (blue dashed lines) for series observed at 3 p.m. at Fort St. John Airport. }
% \end{figure*}

\vspace{0.2 in}

\clearpage
\section{Supplementary Tables}
~
\begin{table}[H]
\caption{Parameter estimates (95\% confidence intervals in parentheses) for the marginal mean model in Equation~(2) without changepoint and with the estimated changepoint, based on daily sky-cloudiness conditions observed at 9 a.m. and 3 p.m., with significant parameters from Wald inference bolded.}\label{tab6}
\centering
\setlength{\tabcolsep}{0pt}\fontsize{8}{12} \selectfont
\begin{tabular*}{\textwidth}{@{\extracolsep{\fill}}crrrr}
\toprule
    & \multicolumn{2}{c}{\multirow{1}{*}{Daily series observed at 9 a.m.}} & \multicolumn{2}{c}{\multirow{1}{*}{Daily series observed at 3 p.m.}} \\
    \midrule
        \multicolumn{1}{c}{\multirow{1}{*}{Parameter}}
            & \multicolumn{1}{c}{\multirow{2}{*}{Without changepoint}} 
                & \multicolumn{1}{c}{\multirow{2}{*}{With changepoint}} 
                    & \multicolumn{1}{c}{\multirow{2}{*}{Without changepoint}} 
                        & \multicolumn{1}{c}{\multirow{2}{*}{With changepoint}}\\
         \multicolumn{1}{c}{\multirow{1}{*}{Estimates}}
            & %\multicolumn{1}{c}{\multirow{1}{*}{without changepoint}}  
                & %\multicolumn{1}{c}{\multirow{1}{*}{with changepoint}}  
                    & %\multicolumn{1}{c}{\multirow{1}{*}{without changepoint}}
                        & %\multicolumn{1}{c}{\multirow{1}{*}{with changepoint}}  
                        \\
        \midrule
        \multicolumn{5}{c}{Marginal Mean Model} \\
        \midrule
        $\alpha_{1}$ & \textbf{-3.3366 (-3.5346, -3.1386)} & \textbf{-3.2191 (-3.4179, -3.0204)} & \textbf{-3.6966 (-3.9316, -3.4616)} & \textbf{-3.5341 (-3.771, -3.2972)} \\
$\alpha_{2}$ & \textbf{-1.2448 (-1.3439, -1.1457)} & \textbf{-1.1834 (-1.2891, -1.0778)} & \textbf{-1.4945 (-1.5990, -1.3900)} & \textbf{-1.3741 (-1.4848, -1.2634)} \\
$\alpha_{3}$ & \textbf{-0.8811 (-0.9732, -0.7890)} & \textbf{-0.8521 (-0.9511, -0.753)} & \textbf{-0.9406 (-1.0329, -0.8484)} & \textbf{-0.8612 (-0.9607, -0.7617)} \\
$\alpha_{4}$ & \textbf{1.0856 (0.9813, 1.1899)} & \textbf{1.0323 (0.9205, 1.1441)} & \textbf{1.1936 (1.0879, 1.2993)} & \textbf{1.1478 (1.0336, 1.262)} \\
$B_{1}$ & \textbf{-0.1741 (-0.2845, -0.0636)} & \textbf{-0.1830 (-0.2926, -0.0735)} & \textbf{0.4767 (0.3411, 0.6122)} & \textbf{0.4879 (0.3529, 0.6229)} \\
$B_{2}$ & \textbf{-0.2418 (-0.3108, -0.1729)} & \textbf{-0.2409 (-0.3097, -0.1720)} & \textbf{0.0942 (0.0219, 0.1666)} & \textbf{0.0975 (0.0255, 0.1695)} \\
$B_{3}$ & \textbf{-0.2468 (-0.3112, -0.1823)} & \textbf{-0.2482 (-0.3128, -0.1837)} & \textbf{-0.0713 (-0.1365, -0.0061)} & \textbf{-0.0686 (-0.1338, -0.0035)} \\
$B_{4}$ & \textbf{-0.4015 (-0.4754, -0.3276)} & \textbf{-0.3990 (-0.4728, -0.3251)} & \textbf{-0.4337 (-0.5093, -0.3582)} & \textbf{-0.4315 (-0.5070, -0.3561)} \\
$D_{1}$ & \textbf{0.1667 (0.0540, 0.2794)} & \textbf{0.1681 (0.0565, 0.2797)} & 0.0394 (-0.0903, 0.1691) & 0.0328 (-0.0958, 0.1614) \\
$D_{2}$ & 0.0379 (-0.0335, 0.1092) & 0.0388 (-0.0324, 0.1100) & -0.0380 (-0.1105, 0.0345) & -0.0457 (-0.1178, 0.0264) \\
$D_{3}$ & 0.0360 (-0.0305, 0.1026) & 0.0366 (-0.0299, 0.1031) & -0.0226 (-0.0880, 0.0428) & -0.0277 (-0.0930, 0.0377) \\
$D_{4}$ & -0.0456 (-0.1179, 0.0267) & -0.0401 (-0.1124, 0.0322) & -0.0663 (-0.1395, 0.0069) & -0.0588 (-0.1320, 0.0144) \\
$\beta_{1}$ & \textbf{1.5302 (1.2294, 1.831)} & -0.1535 (-0.6718, 0.3648) & \textbf{1.6231 (1.2711, 1.9751)} & -0.2379 (-0.8665, 0.3907) \\
$\beta_{2}$ & \textbf{0.3400 (0.1692, 0.5109)} & -0.1274 (-0.4654, 0.2106) & \textbf{0.6226 (0.4467, 0.7985)} & -0.3058 (-0.6552, 0.0436) \\
$\beta_{3}$ & \textbf{0.2996 (0.1395, 0.4597)} & 0.0846 (-0.2327, 0.4020) & \textbf{0.4279 (0.2683, 0.5875)} & -0.1379 (-0.4557, 0.1798) \\
$\beta_{4}$ & \textbf{-0.1981 (-0.3773, -0.0189)} & 0.1944 (-0.1580, 0.5467) & 0.0117 (-0.1715, 0.1948) & 0.3412 (-0.0161, 0.6985) \\
$\Delta_{1}$ & \textbf{-} & \textbf{1.2513 (0.9352, 1.5674)} & \textbf{-} & \textbf{1.3408 (0.9621, 1.7195)} \\
$\Delta_{2}$ & \textbf{-} & \textbf{0.3180 (0.1193, 0.5166)} & \textbf{-} & \textbf{0.6337 (0.4269, 0.8405)} \\
$\Delta_{3}$ & \textbf{-} & 0.1460 (-0.0395, 0.3314) & \textbf{-} & \textbf{0.3807 (0.1949, 0.5665)} \\
$\Delta_{4}$ & \textbf{-} & \textbf{-0.2626 (-0.4665, -0.0587)} & \textbf{-} & \textbf{-0.2201 (-0.4246, -0.0156)} \\
        \midrule
        \multicolumn{5}{c}{Dependence Model} \\
        \midrule
        $\xi_{11}$ & \textbf{1.9276 (1.6455, 2.2096)} & \textbf{1.8997 (1.6188, 2.1806)} & \textbf{2.3302 (2.0058, 2.6546)} & \textbf{2.3083 (1.9844, 2.6322)} \\
$\xi_{12}$ & \textbf{1.3352 (1.0832, 1.5872)} & \textbf{1.3564 (1.1037, 1.6092)} & \textbf{1.2786 (0.9877, 1.5696)} & \textbf{1.2857 (0.9938, 1.5775)} \\
$\xi_{13}$ & \textbf{1.3623 (1.0271, 1.6976)} & \textbf{1.3927 (1.0565, 1.7290)} & \textbf{1.1932 (0.8245, 1.5618)} & \textbf{1.2269 (0.8568, 1.5969)} \\
$\xi_{14}$ & \textbf{0.8366 (0.6129, 1.0604)} & \textbf{0.8549 (0.6306, 1.0791)} & \textbf{0.6781 (0.4170, 0.9393)} & \textbf{0.7034 (0.4409, 0.9658)} \\
$\xi_{21}$ & \textbf{1.2750 (1.0333, 1.5166)} & \textbf{1.2941 (1.0519, 1.5363)} & \textbf{1.2220 (0.9385, 1.5055)} & \textbf{1.2279 (0.9441, 1.5118)} \\
$\xi_{22}$ & \textbf{1.1023 (0.9188, 1.2859)} & \textbf{1.0908 (0.9070, 1.2747)} & \textbf{1.0771 (0.8870, 1.2671)} & \textbf{1.0702 (0.8801, 1.2603)} \\
$\xi_{23}$ & \textbf{1.0281 (0.7688, 1.2874)} & \textbf{1.0153 (0.7553, 1.2753)} & \textbf{0.7950 (0.5470, 1.0431)} & \textbf{0.7925 (0.5443, 1.0408)} \\
$\xi_{24}$ & \textbf{0.8486 (0.6977, 0.9994)} & \textbf{0.8417 (0.6906, 0.9928)} & \textbf{0.6373 (0.4805, 0.7942)} & \textbf{0.6376 (0.4805, 0.7946)} \\
$\xi_{31}$ & \textbf{0.8677 (0.5130, 1.2224)} & \textbf{0.8933 (0.5375, 1.2490)} & \textbf{0.7682 (0.3876, 1.1489)} & \textbf{0.7992 (0.4184, 1.1800)} \\
$\xi_{32}$ & \textbf{0.9608 (0.7037, 1.2179)} & \textbf{0.9482 (0.6910, 1.2054)} & \textbf{0.7717 (0.5293, 1.0141)} & \textbf{0.7692 (0.5267, 1.0118)} \\
$\xi_{33}$ & \textbf{1.1052 (0.7647, 1.4458)} & \textbf{1.0846 (0.7428, 1.4263)} & \textbf{0.7866 (0.4924, 1.0808)} & \textbf{0.7640 (0.4690, 1.0591)} \\
$\xi_{34}$ & \textbf{0.8795 (0.6694, 1.0896)} & \textbf{0.8653 (0.6553, 1.0753)} & \textbf{0.5723 (0.3773, 0.7673)} & \textbf{0.5546 (0.3591, 0.7501)} \\
$\xi_{41}$ & \textbf{0.5384 (0.3221, 0.7547)} & \textbf{0.5541 (0.3372, 0.7709)} & \textbf{0.5516 (0.2959, 0.8073)} & \textbf{0.5743 (0.3184, 0.8302)} \\
$\xi_{42}$ & \textbf{0.7347 (0.5862, 0.8832)} & \textbf{0.7280 (0.5793, 0.8766)} & \textbf{0.6396 (0.4823, 0.7968)} & \textbf{0.6396 (0.4823, 0.797)} \\
$\xi_{43}$ & \textbf{0.8075 (0.5945, 1.0205)} & \textbf{0.7932 (0.5801, 1.0062)} & \textbf{0.7301 (0.5300, 0.9303)} & \textbf{0.7126 (0.5121, 0.9132)} \\
$\xi_{44}$ & \textbf{0.7094 (0.5952, 0.8237)} & \textbf{0.7004 (0.5861, 0.8148)} & \textbf{0.6537 (0.5344, 0.7731)} & \textbf{0.6408 (0.5212, 0.7604)} \\
\bottomrule
\end{tabular*}
\end{table}

\end{supplement}
% \begin{supplement}
% \stitle{Data and R code}
% \sdescription{Short description of Supplement B.}
% \end{supplement}

%%  imsart-nameyear.bst  will be used to                   %%
%%  MR numbers will be added by VTeX.                      %%
%%                                                         %%
%%  Use \cite{...} to cite references in text.             %%
%%                                                         %%
%%%%%%%%%%%%%%%%%%%%%%%%%%%%%%%%%%%%%%%%%%%%%%%%%%%%%%%%%%%%%

%% if your bibliography is in bibtex format, uncomment commands:
\bibliographystyle{imsart-nameyear} % Style BST file
\bibliography{bibliography}       % Bibliography file (usually '*.bib')

%% or include bibliography directly:
%% \begin{thebibliography}{4}
%%
% \bibitem[\protect\citeauthoryear{Billingsley}{1999}]{r1}
% \textsc{Billingsley, P.} (1999). \textit{Convergence of
% Probability Measures}, 2nd ed.
% Wiley, New York.

% \bibitem[\protect\citeauthoryear{Bourbaki}{1966}]{r2}
% \textsc{Bourbaki, N.}  (1966). \textit{General Topology}  \textbf{1}.
% Addison--Wesley, Reading, MA.

% \bibitem[\protect\citeauthoryear{Ethier and Kurtz}{1985}]{r3}
% \textsc{Ethier, S. N.} and \textsc{Kurtz, T. G.} (1985).
% \textit{Markov Processes: Characterization and Convergence}.
% Wiley, New York.

% \bibitem[\protect\citeauthoryear{Prokhorov}{1956}]{r4}
% \textsc{Prokhorov, Yu.} (1956).
% Convergence of random processes and limit theorems in probability
% theory. \textit{Theory  Probab.  Appl.}
% \textbf{1} 157--214.
% \end{thebibliography}

\end{document}